\documentclass[%
 reprint,
 superscriptaddress,
 onecolumn,
 amsmath,amssymb,
 aps,
]{revtex4-2}

\usepackage[utf8]{inputenc}
\usepackage[english]{babel}

\usepackage{graphicx}% Include figure files
  \graphicspath{{figures/}}
\usepackage{dcolumn}% Align table columns on decimal point
\usepackage{bm}% bold math

\usepackage{subfiles}  % use subfiles
\usepackage{xcolor} % colored text

\usepackage{bibentry}
\nobibliography*

\begin{document}
\title{Intelligent metaphotonics empowered by machine learning}

\author{Sergey Krasikov}%
  \email{s.krasikov@metalab.ifmo.ru}
  \affiliation{School of Physics and Engineering, ITMO University, St. Petersburg 197101, Russia}%
    \affiliation{Nonlinear Physics Center, Research School of Physics, Australian National University, Canberra ACT 2601, Australia}%
\author{Aaron Tranter}%
  \affiliation{Centre for Quantum Computation and Communication Technology, Department of Quantum Science, Research School of Physics, The Australian National University, Canberra, ACT 2601, Australia}%
\author{Andrey Bogdanov}%
  \affiliation{School of Physics and Engineering, ITMO University, St. Petersburg 197101, Russia}%
\author{Yuri Kivshar}%
 \email{yuri.kivshar@anu.edu.au}
  %\email{Corresponding author: Yuri Kivshar: yuri.kivshar@anu.edu.au}
    \affiliation{School of Physics and Engineering, ITMO University, St. Petersburg 197101, Russia}%
  \affiliation{Nonlinear Physics Center, Research School of Physics, Australian National University, Canberra ACT 2601, Australia}%

\date{\today}

\begin{abstract}
In the recent years, we observe a dramatic boost of research in photonics empowered by the concepts of {\it machine learning} and {\it artificial intelligence}. The corresponding photonic systems, to which this new methodology is applied, can range from traditional optical waveguides to nanoantennas and metasurfaces, and these novel approaches underpin the fundamental principles of light-matter interaction developed for a smart design of intelligent photonic devices.  Concepts and approaches of artificial intelligence and machine learning penetrate rapidly into the fundamental physics of light, and they provide effective tools for the study of the field of  {\it metaphotonics} driven by optically-induced electric and magnetic resonances. Here, we introduce this new field with its application to metaphotonics and also present a summary of the basic concepts of machine learning with some specific examples developed and demonstrated for metasystems and metasurfaces. 
\end{abstract}

\maketitle

\section{Introduction}

As demonstrated in the recent years, artificial intelligence (AI) is quickly becoming a ubiquitous concept in various physical sciences~\cite{radovic2018machine,schmidt2019recent,brown2020machine,carrasquilla2020machine,bedolla2020machine,campbell2021explosion}. Specifically, optics and photonics provide a variety of physical applications that can be studied and enhanced using these novel approaches. Merging AI with photonics is particularly advantageous as Maxwell’s equations fundamentally describe very precisely a wide range of experimentally observed phenomena. Thus they can be employed  directly to generate vast amounts of training data required for implementing various AI algorithms.  Also, by employing AI in optical physics, we may expect to create efficient photonic quantum machines that eventually may merge with much broader biological intelligence. Thus, a remarkably powerful AI methodology that complements many conventional analytical and numeric techniques find many important applications and demonstrations in photonics. In particular, such approaches are used for inverse design, optimization, big-data processing, underpinning the rapid development of precise photonic technologies. 

{\it Metaphotonics} is a new and rapidly developing research direction in subwavelength photonics~\cite{koshelev2021dielectric}. It is inspired by the physics of metamaterials where the electromagnetic response is associated with the magnetic dipole resonances and optical magnetism originating from the resonant dielectric nanostructures with high refractive index.  The concept of all-dielectric resonant nanophotonics is driven by the idea to employ subwavelength dielectric Mie-resonant nanoparticles as ``meta-atoms'' for creating highly efficient optical metasurfaces and metadevices. These are defined as devices having unique functionalities derived from structuring of functional matter on the subwavelength scale~\cite{zheludev2012metamaterials}, often termed as ``meta-optics'' for emphasizing the importance of optically-induced magnetic response of their artificial subwavelength-patterned structures. Because of the unique optical resonances and their various combinations for interference effects accompanied by strong localization of the electromagnetic fields, high-index nanoscale structures are expected to complement or even replace different plasmonic components in a range of potential applications. Moreover, many concepts which had been developed for plasmonic structures, but fell short of their potential due to strong losses of metals at optical frequencies, can now be realized based on low-loss dielectric structures.

High-index dielectric resonators can support both electric and magnetic Mie-type resonances which can be tailored by the nanoparticle geometry. Mie-resonant silicon nanoparticles have recently received considerable attention for applications in nanophotonics and metamaterials~\cite{kruk2017functional} including optical nanoantennas, wavefront-shaping metasurfaces, and nonlinear frequency generation. Importantly, the simultaneous excitation of strong electric and magnetic Mie-type dipole and multipole resonances can result in constructive or destructive interferences with unusual spatial scattering characteristics, and it may also lead to the resonant enhancement of magnetic fields in dielectric structures that could bring many novel effects in both linear and nonlinear regimes.

Here, we discuss the current progress in the recently emerged fields of machine learning and artificial intelligence focusing on their applications to  metaphotonics driven by optically-induced electric and magnetic resonances. More specifically, we describe how these novel ideas can be applied to study optical nanoantennas and metasurfaces and improve their functionalities. We also present a summary of the basic concepts of machine learning with some specific examples developed and demonstrated for a variety of metasystems and metasurfaces. 

This Perspective is organized as follows. In Sec.~\ref{sec:ML_concepts} we broadly discuss the basic concepts of machine learning irrespective of specific applications. We introduce the concepts of deep learning and machine learning and show how they are linked to artificial neutral networks and artificial intelligence concepts.  Next, in Sec.~\ref{sec:metasystems} we summarize briefly the main ideas of metaphotonics and describe the main parameters and properties which are required usually for the optimisation. In Sec.~\ref{sec:nanoantennas} we consider several examples of advanced nanoantennas and their optimisation with the help of machine learning. All Secs.~\ref{sec:metasurfaces}~--~\ref{sec:self_adapting} discuss various aspects of functional and self-adaptive metasurfaces and their applications to sensing. Finally, Sec.~\ref{sec:perspective} concludes the paper with perspectives and outlook.

\section{Basic concepts of machine learning}
\label{sec:ML_concepts}

In this Perspective article, we focus on phenomena and applications which can be reached via AI technologies, and we do not discuss particular methods in detail. Still, here we provide a brief description of machine learning (ML) algorithms in order to outline the general ideas of those methods suitable for applications.

While AI is a broad field closely connected with the development of systems mimicking intelligence, ML provides a set of data-driven algorithms for learning ability of AI~\cite{mohri2018foundations}.
The goal of ML algorithms is to find a sequence of programmatic transformations connecting the input and output data. This could constitute a mathematical model that connects physical parameters to an observed phenomena~\cite{mehta2019highbias,roscher2020explainable,tanaka2021deep}. In another case it could be the use of a heuristic to classify labelled data. Considering an example from physics, this might be transformations which have to be applied to geometric parameters of a unit cell of a metagrating to obtain a corresponding reflection spectrum. Generally, provided a good source of data, the more input data the algorithm receives, the more accurate a result will be. It should be highlighted that ML algorithms can work with data itself and find correlations within it without any understanding of physical, mathematical, or other meaning. However, in some cases it is possible to provide physical intuition or to constrain models to obey physical models such as, for example, the Navier-Stokes equations~\cite{miyanawala_efficient_2017, lucor_physics-aware_2021} or Maxwell's equations~\cite{lim2021maxwellnet}. Existing ML methods span a range of tasks, including but not limited to, classification, regression, clustering, anomaly detection and structured prediction. While ML has demonstrable efficacy for a number of tasks there remains significant overhead involved in the application of such techniques. This usually involves a number of tasks such as tuning and design of the models in question, with the largest task generally being the feature extraction process which often requires domain specific expertise. 

\begin{figure}[htbp!]
  \centering
  \includegraphics[width=\linewidth]{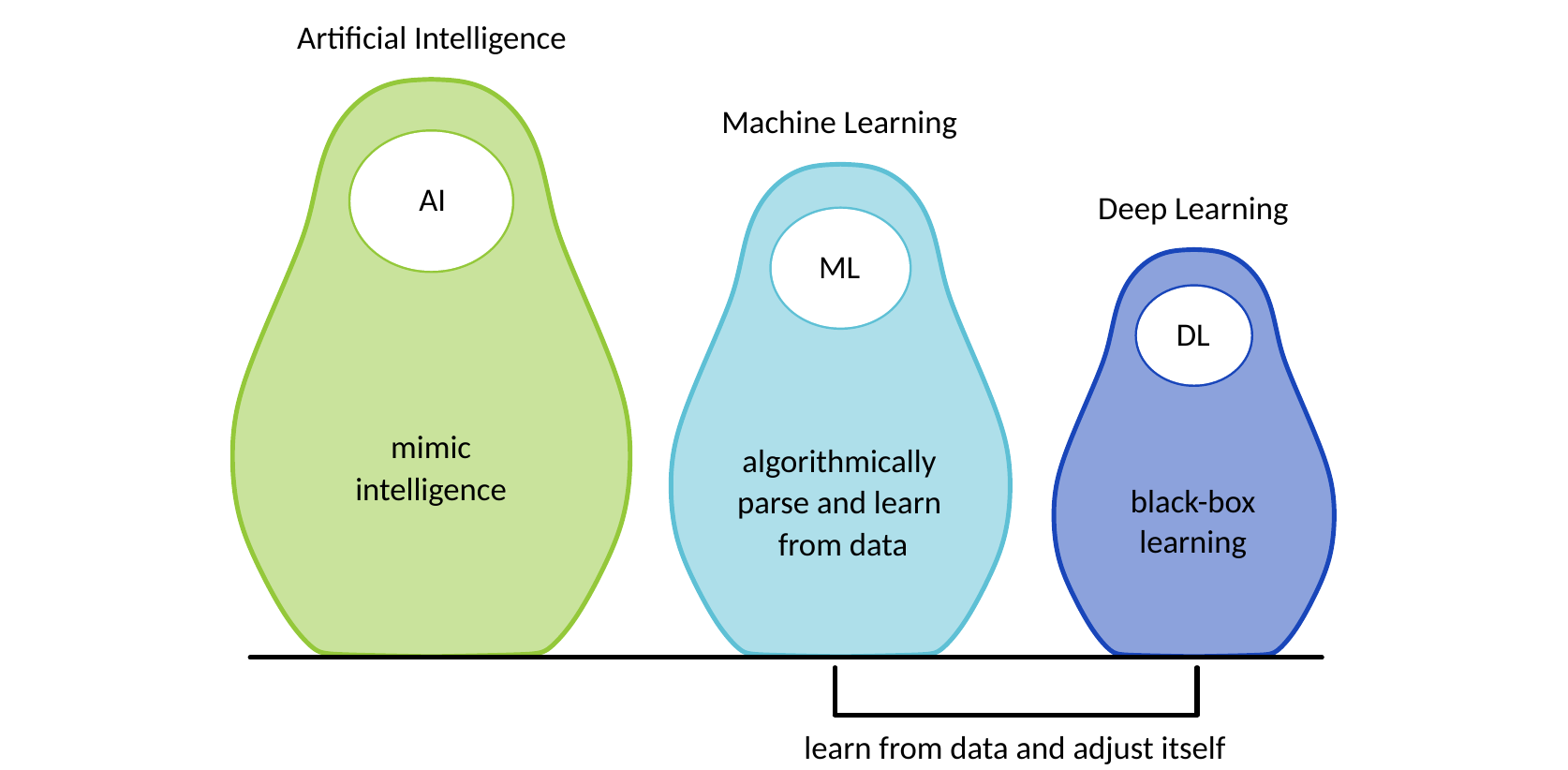}
  \caption{{\bf Links of different ML-based concepts.} Artificial intelligence (AI) is a part of computer science dedicated to development of ways to mimic general intelligence. Machine learning (ML) is a subset of AI, these are data-driven algorithms which learn from experience and have the capacity to improve their performance over time and adapt to new data. ML algorithms are varied in their approach, however deep learning (DL) is a subset of ML solely based on layered structures referred to as artificial neural networks. The main feature of DL is the capability to efficiently process raw unstructured data and automatically determine its features while classical ML involves the processing of data in a manner predefined by a human operator.}
  \label{fig:AI}
\end{figure}

Deep learning (DL) is a subclass of ML (see Fig.~\ref{fig:AI}) solely based on layered structures referred to as artificial neural networks (ANNs). ANNs derive their name from the neural structures found in biological entities. This structure is emulated mathematically as a node, referred to as a neuron, which may contain many input and output connections with associated weightings. These neurons have a non-linear activation function which is a function that serves to map the inputs to an output and provide the switching behaviour seen in biological neurons~\cite{goodfellow_deep_2016}. These neurons are stacked into layers which are then connected to subsequent layers. The utility of this is to construct a sufficiently deep neural network such that any arbitrary function may be approximated~\cite{hornik1989multilayer}. ANNs thus may be considered as a mapping from some input space to some output space which can be arbitrarily defined. ANNs can also have complicated structures for different tasks. The inclusion of convolutional layers have demonstrated great success in image processing tasks~\cite{simonyan_very_2015} and auto-regressive models similarly with natural language processing~\cite{brown_language_2020} and sequentially organised data~\cite{oord_pixel_2016}. A key feature of these models and DL in general is that the design process does not require the same domain specific knowledge as other classical ML methods. Instead the features of a data set are learnt automatically to best facilitate the desired mapping from input to output. To perform a given task DL models must undergo a process known as training.  Training a model requires the introduction of the so-called loss function, which provides feedback on the difference between the real output (or ``truth'') and the output predicted by the network for the same given input data. Training the ANN aims to minimize the loss function by adjustment of the weight values at each layer. This tasks is generally achieved via some form of stochastic gradient descent which may be done efficiently over the network with backward propagation~\cite{rumelhart_learning_1986}. The training is termed complete when the model is capable of predicting the output with some desired accuracy.

After the training, the ANN is able to accurately map the input to the desired output, implying that it has ``learned'' the necessary mapping from the given data. It should be noted that this mapping is not necessarily unique and differences in training procedure can cause a failure to converge. It is generally considered good practice to divide a dataset into training and validation sets. Due to the expressivity of high dimensional networks it is possible for over-fitting to occur which must be mitigated. Over-fitting can be monitored via the loss function applied to the validation set. Once the model is trained and over-fitting is mitigated the model can be said to be capable of generalising for an out of sample input. This of course assumes that the training data provided is representative of the problem that one is trying to learn. It is important to note that in general ML models are interpolative by construction, thus extrapolative intuition (not to be confused with generalisation) is still a difficult problem that is indeed difficult for human operators as well. 

\begin{figure}[htbp!]
  \centering
  \includegraphics[width=\linewidth]{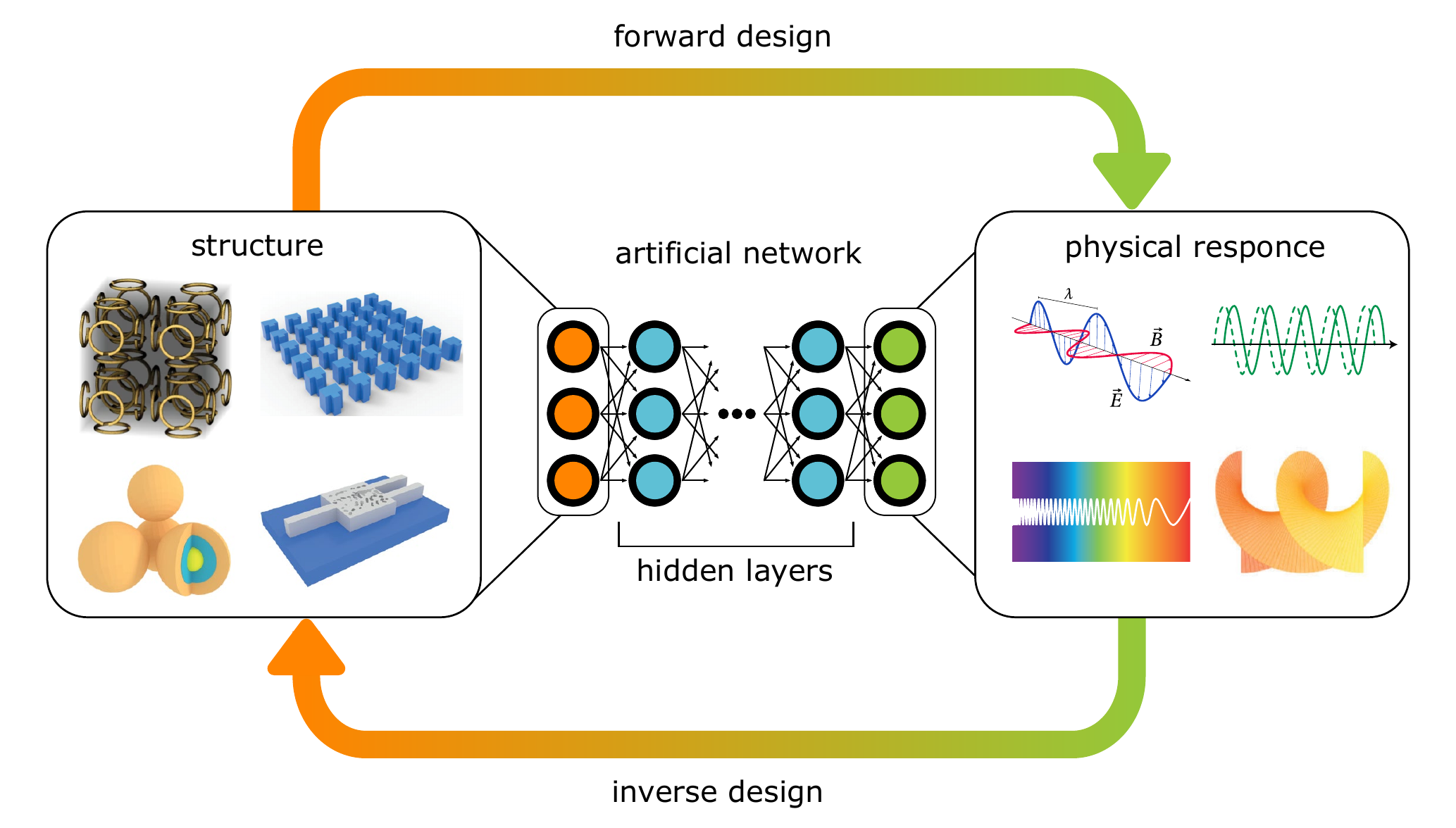}
  \caption{{\bf Inverse and forward designs based on DL techniques}. An artificial neural network is fed parameters of a structure and the corresponding physical response obtained with conventional numerical techniques or from an experiment. Structure parameters may come in a form of numerical arrays containing geometric or material parameters, or in a form of images or sets of pixels. During the learning procedure the neural network determines the appropriate mapping to account for the relation between the parameters and the responses. Then, the network is able to predict a structures response which was not present in the learning set of data – this is {\it a forward design}, which is a prediction of a physical response for a given structure. Swapping the input and output data and applying the same training procedure, the network may provide a reverse function, predicting parameters of a structure which allows achieving a given physical response, called {\it an inverse design.}}
  \label{fig:design}
\end{figure}

One of the major drawbacks of ML is the amount of data needed for the training procedure, which is typically dictated by the complexity of the problem. For DL increased training data also tends to mitigate over-fitting to a certain extent and thus can be even more important. The collection of the data as well as training may take an enormous time. Hence, the use of DL is usually justified when the same task has to be solved many times for slightly different set of parameters or the task of building a feature extractor is intractable. 

Concerning photonics, ML was so-far incorporated mainly as a tool for forward and inverse design procedures (see Fig.~\ref{fig:design}). Forward design is associated with prediction of a physical response (scattering spectra, polarization, etc.) for a given structure. This task can be solved via variety of different tools, such as T-matrix calculations, full-wave simulations and others, which are basically aimed to solve Maxwell's equations in some form. Though ML may become an alternative method for the same purpose, possibly its greatest benefit is associated with inverse design problems --- determination of structure parameters necessary to provide a given response. In this case DL can be used as a basis for the so-called surrogate modelling providing not a simulation but rather its data-driven approximation.

To illustrate the above ideas, we consider an example of a metaphotonic system which can be designed with the help of an ANN. Let the aim be to predict a scattering spectrum for a given radius of a uniform spherical particle. In this case the input is the radius, and the output is the value of a scattering cross-section (SCS) at some particular frequency. The training set consists of pairs [radius, SCS] and the network finds the mapping between these two values, meaning that it finds a sequence of mathematical transformations which converts a given radius to a known SCS value. In this case, the transformations are not constrained to have anything in common with any physical equation, theory, or model.

When the sequence is defined, the network can predict the unknown value of SCS for a radius which was not given during the training.  Importantly, the reverse process is possible, and inputs can be swapped with outputs so the network will estimate parameters required for a given SCS value. No additional training procedures or any changes of the algorithm are needed --- when the connections are determined they work in both ways. Therefore, the network can be trained with the data acquired by conventional solvers and then it is capable to solve the inverse task to find parameters for a desired physical response. However, it should be mentioned that for real-world tasks the procedure may be not that straightforward and to achieve high accuracy special architectures of ANNs and algorithms should be considered. Nevertheless, accuracy of the predictions may approach unity when provided enough data, with the speed of estimation suitable to many real-time tasks making DL a promising tool for design of photonic devices. 

The same logic can be applied for other classes of problems, not necessary related to scattering. However, in this Perspective we  describe not the algorithms themselves but some results which can be achieved with their help. Thus, below we refer to DL algorithms as black-boxes with unspecified structures and functional mappings between the physical parameters.

\section{Metasystems and Metasurfaces} 
\label{sec:metasystems}

Electromagnetic metamaterials were suggested as artificial structures composed of subwavelength elements, and initially they were driven by curiosity such as negative refraction and super-lens. Some years later, metamaterials created a paradigm for engineering electromagnetic space with the help of transformation optics. More recently, research on metamaterials evolved into the study of metasystems and metasurfaces as basic components of the so-called {\it metadevices} defined as optical devices having unique and useful functionalities realized by structuring of functional matter on the subwavelength scale~\cite{zheludev2012metamaterials}. It is expected that future technologies will involve a high level of photonic integration achieved by embedding the data-processing and waveguiding functionalities at the material’s level. Thus, an important task is to study and optimize materials building blocks such as nanoantennas and metasurfaces. 

{\it Plasmonic} and {\it dielectric nanoantennas} supporting resonances represent a novel type of building blocks of metamaterials for generating, manipulating, and modulating light~\cite{li2021directional, koshelev2021dielectric}. By combing both electric and magnetic modes, one can not only modify far-field radiation patterns but also localize the electromagnetic energy in open resonators by employing the physics of bound states in the continuum to achieve destructive interference of leaky modes~\cite{rybin2017high}.

{\it Metasurfaces} are created by artificial subwavelength elements with small thickness which provides novel capability to manipulate electromagnetic waves~\cite{kildishev2013planar}. Well before the exploration of metasurfaces, tailoring the light scattering with planar optical structures has been majorly pursued with diffractive optical elements~\cite{lalanne2017metalenses}. However, the concept of metasurfaces provides much broader and deeper insights and useful tools for complete control of light. Metasurfaces are characterized by reduced dimensionality, and usually they consist of arrays of optical resonators with spatially varying geometric parameters and subwavelength separation. In contrast to conventional optical components that achieve wavefront engineering by phase accumulation through light propagation in a medium, metasurface provides new degrees of freedom to control the phase, amplitude, and polarization of light waves with subwavelength resolution, as well as to accomplish wavefront shaping within a distance much less than the wavelength of light. The outstanding optical properties of dielectric metasurfaces drive the development of ultra-thin optical elements and devices, whether showing novel optical phenomena or new functionalities outperforming their traditional bulky counterparts~\cite{kruk2017functional,hsiao2017fundamentals,qiu2021quo}. Metasurfaces consist of carefully arranged ``unit cells'' or ``meta-atoms'' with subwavelength structures. The optical response (phase, amplitude, and polarisation) of the meta-atom changes with its geometry (height, width, material, etc). The meta-atoms are arranged into arrays to provide specific variations of parameters, depending on required functionalities. Meta-atoms can operate as subwavelength resonators supporting multipolar Mie resonances~\cite{kruk2017functional}, or they can contribute to averaged parameters like metamaterials~\cite{hsiao2017fundamentals}.

The concept of optical metasurfaces have been applied to demonstrate many exotic optical phenomena and various useful planar optical devices. Many of these metasurface-based applications are potentially very promising alternatives to replace conventional optical elements and devices, as they largely benefit from ultra-thin, lightweight, and ultracompact properties, provide the possibility of overcoming several limitations suffered by their traditional counterparts, and can demonstrate versatile novel functionalities. Metasurfaces were suggested for an efficient control of light-matter interaction with subwavelength resonant structures, and they have been explored widely in the recent years for creating transformational flat-optics devices.

Plasmonic metastructures suffer from significant losses and show low efficiency, but all-dielectric structures can readily combine electric and magnetic Mie resonances and control efficiently optical properties such as amplitude, phase, polarization, chirality, and anisotropy~\cite{koshelev2021dielectric}. The control of all such parameters require a careful optimisation depending on the problems where the metasurfaces are used. Many such properties are driven by local electromagnetic resonances such as Mie-type scattering, bound states in the continuum, Fano resonances, and anapole resonances. The recent research frontiers in dielectric metasurfaces include wavefront-shaping, metalenses, multifunctional and computational approaches, with the main strategies to realize the dynamic tuning of dielectric metasurfaces.

Importantly, recent advances in nanofabrication technologies bring low-cost, large-area and mass productive approaches and capabilities for the development of various types of metasystems and metasurfaces, and the methods are gradually becoming mature. It is expected that flat-optics components based on dielectric metasurfaces will appear in our daily life very soon bringing complexity of optical components and novel functionalities~\cite{chen2021will}.

\section{Advanced nanoantennas}
\label{sec:nanoantennas}

One of the first illustrative examples for application of the ML techniques to metaphotonics is a design of nanoantennas as elementary units of metasystems meta-atoms.  Engineering of a scattering response is often realized with core-shell structures via tuning of thickness and dielectric permittivity of the layers. In this case, a design of nanoantennas may be time consuming due to the overwhelming number of possible combinations of the nanoantenna parameters. Even accounting for manufacturing limitations, the number of possible materials and layer thicknesses can go far beyond thousands of values. Without advanced optimization techniques they should be iterated to fit a desired scattering spectrum, which implies a large number of calculations. In this case, the application of DL techniques can become very useful and productive.

\begin{figure}[htbp!]
  \centering
  \includegraphics[width=\linewidth]{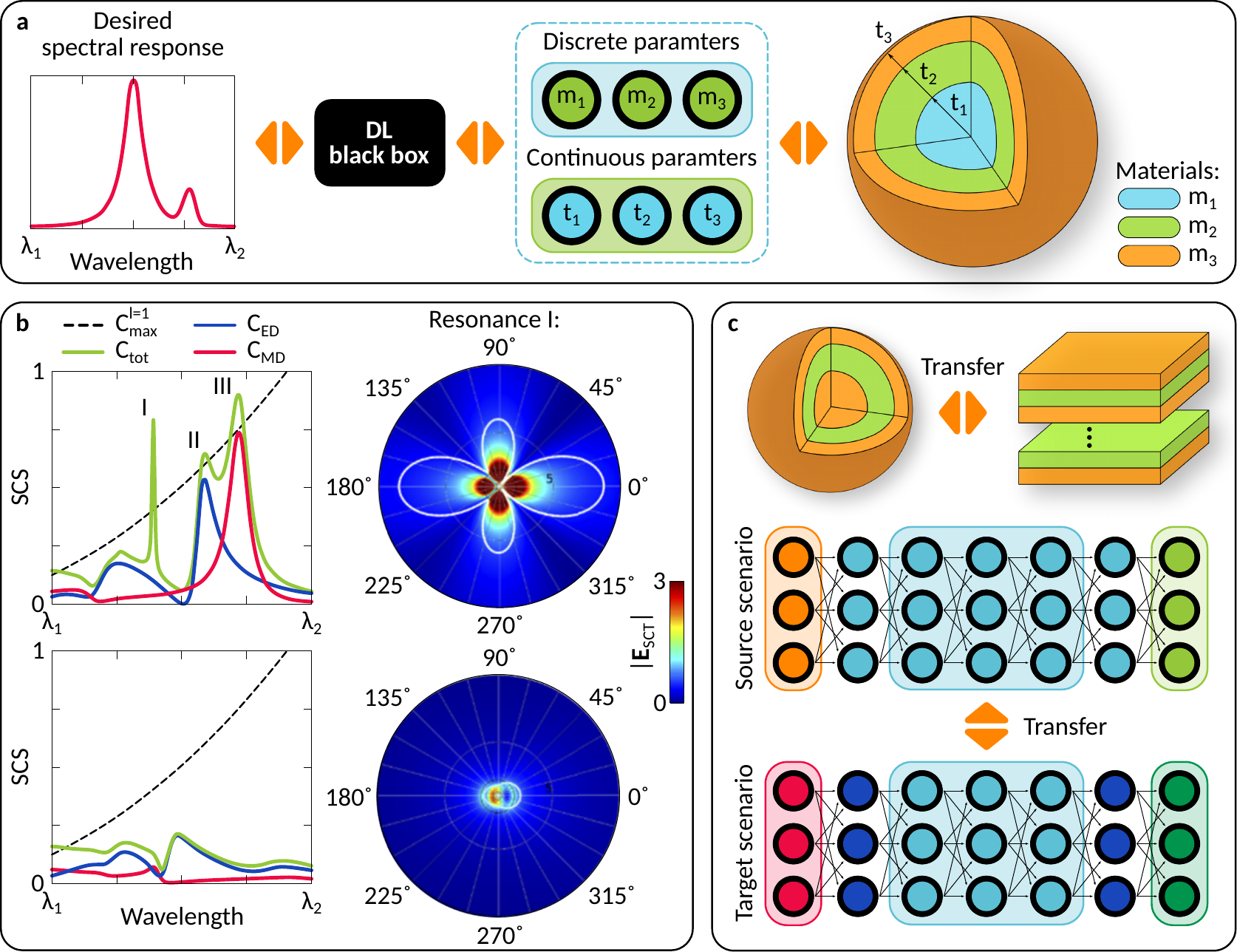}
  \caption{{\bf ML-empowered designs of multilayer nanoantennas.} (a) Schematic of the design procedure based on the DL approaches. A DL algorithm (a black box) connects physical response with parameters of a structure. For example, hand-drawn scattering spectra can be processed with the black-box in order to define materials and thicknesses of the layers needed to achieve target spectra (Based on the results from  Ref.~\cite{so2019simultaneous}). (b) Demonstration of invisibility-to-superscattering transition of a multilayer sphere made of phase-change materials. Materials and thicknesses for this case are found similar to the scheme in (a). (Adopted from Ref.~\cite{luo2021deeplearningenabled}). (c) Transfer learning process. Layers of one trained ANN can be merged with another ANN in order to provide a training procedure for another type of the structure. This might be used for a design of multilayer films  or multilayer sphere with using the ANN approach (Concept originates from the results of Ref.~\cite{qu2019migrating,qiu2021nanophotonic}).}
  \label{fig:antennas}
\end{figure}

The general idea of a design procedure based on the DL approach is  shown schematically in Fig.~\ref{fig:antennas}(a). One of the first results in this field was presented in Ref.~\cite{peurifoy2018nanophotonic} where the network was trained to determine a scattering spectrum of a layered spherical nanoparticle with fixed material parameters. The inputs of the network are the thicknesses of the layers ($30$ --~$70$~nm) and the outputs are the scattering spectrum samples in the range of wavelengths between $400$~nm and $800$~nm. After the training, the network is able to generate a scattering spectrum for a given set of parameters within a second and with a high precision (mean relative error is below $1.5$\%). Importantly, the re-trained network is used to provide an inverse design such that for a given spectra it gives a suitable set of thicknesses. Moreover, the network is used as a tool for optimizing the spectral features in narrow and broadband regions of the spectra. Similar tasks have also been considered in  Refs.~\cite{hu2019robust,qiu2020inverse}, where more advanced DL techniques have been applied. 

The DL procedure was further generalized to adjust not only thicknesses but also materials of a three-layered sphere via the application of neural network~\cite{so2019simultaneous}. In this case, the inputs are electric and magnetic dipole responses while the output is the set of materials and thicknesses. To avoid arbitrary refractive indices which cannot be achieved in a real situation, the set of possible refractive indices is limited to $7$ values corresponding to typical materials used for nanofabrication (such as Si, SiO$_2$, Ag, etc.). On the one hand, such a strategy significantly limits the possible output parameters, but the computational problem actually become more complicated. In this case, two tasks have to be solved simultaneously: a regression problem, to estimate geometric parameters taking continuous values, and classification problem, to determine material properties. Additionally the network must also reconstruct the spectrum from the design parameters. The proposed solution is based on a purpose built hybrid architecture. Two networks are used here: one, to connect optical properties with parameters of the structure, and the other one, to map design parameters to optical responses. Importantly the training of both networks happens simultaneously with a combined loss function that imposes the networks learn forward and inverse design. The proposed DL architecture allows to find and tune electric and magnetic dipole resonances. For example, the algorithm is used to achieve the resonance spectral co-location. Simultaneous tuning of electric and magnetic dipole resonances also allow implementing the first Kerker condition and design even more complex media with negative index of refraction. 

Implementation of ML-based numerical design techniques is not limited to spherical geometries and neural networks. Classical ML algorithm (the so-called Bayesian optimization) was applied for forward design of cylindrical metal-dielectric nanoantenna in order to achieve unidirectional scattering satisfying Kerker or anti-Kerker conditions~\cite{qin2019designing}.
Multipole engineering via DL techniques was also demonstrated, where an ANN was used to determine the far- and near-field response of arbitrary plasmonic and dielectric structures and the internal distribution of fields~\cite{wiecha2020deep}. DL algorithms can be employed to study a variety of problems in metaphotonics including electric and magnetic dipole resonances, Kerker scattering, and destructive interferences leading to the anapole states. 

Advantages of DL-based designs over FDTD simulations of core-shell nanoantennas have been discussed in Ref.~\cite{vahidzadeh2021artificial}. In general, the speed of DL-based forward designs is $100$--$1000$~times faster in comparison with FDTD simulations. At the same time, accuracy of prediction reaches the values about $95\%$. Together with developing ML-supplemented methods for solving both forward and inverse scattering problems~\cite{cao2019hybrid,li2020predicting,guo2021physics,lin2021lowfrequency}, such studies suggest that AI technologies may constitute conventional numerical methods resulting in the development of novel computational tools. For instance, the recent results~\cite{qie2021realtime} suggests real-time web-based tools for designing far-fields of arbitrary-shaped structures. 

Other examples of the scattering problems that can be tackled with DL algorithms include a design of invisible objects. In particular, $5$-layer particle (Ag and SiO$_2$ layers) was inversely designed to achieve extremely low scattering efficiency within an optical frequency spectrum~\cite{sheverdin2020photonic}. Namely, the scattering efficiency was below $10^{-2}$ in the range from $400$~nm to $700$~nm, dropping below $10^{-4}$ between $510$~nm and $550$~nm.
Another demonstration utilizes phase-change materials for realizing invisibility-to-superscattering switching~\cite{luo2021deeplearningenabled}. Developed DL approach allowed to predict required materials and structural parameters to realize simultaneously satisfied conditions for super- and near-zero scattering for two phase states, see Fig.~\ref{fig:antennas}(b). 
The shape of a particle itself may also be a target of DL-based design procedure as for example in Ref.~\cite{blanchard-dionne2021successive} where an all-dielectric shell was designed for optical cloaking.

Design of nanoantennas can also be used as a basis for solving other problems. As was mentioned above, to achieve reasonable accuracy, the application of DL methods requires a large amount of data. At the same time, ANNs are usually implemented for some specific systems,  and a change of the system requires initiation of a new training process even if the general task remains the same. For example, a design of a scattering response of nanoparticles is completely different from the same procedure for metasurfaces though the outputs are quite similar (such as scattering spectra). This is reasonable, since these two systems are characterized by different sets of parameters, and hence there are different mappings between the input and output data. Nevertheless, it is possible to transfer {\it knowledge} gained by an ANN between completely different scenarios, referred to as {\it transfer learning}~\cite{pan2010survey,zhuang2021comprehensive}. This was demonstrated for the transmission of 8-layer films utilizing the results obtained for scattering from 8-layer nanoparticles~\cite{qu2019migrating}. The general idea is presented in Fig.~\ref{fig:antennas}(c), and the procedure requires merging two ANNs meaning that layers of one network (for the scattering problem) can be inserted into the other one (for the transmission problem). As a result, the learning error of the network for the multilayer films is reduced by almost $20\%$ (from $7.1\%$ to $5.7\%$). The proposed technique may be useful for insufficient data allowing to use ANNs already trained for other tasks. Similar idea was developed~\cite{qiu2021nanophotonic} to provide inverse design with finding simultaneous materials and continuous geometric parameters of core-shell particles and multilayer films.

Both shape and size of single-material nanoparticles may also be a target for the inverse design procedure. For instance, DL techniques were implemented to achieve a desired spectral emission~\cite{elzouka2020interpretable} as well as far- and near-field properties~\cite{he2019plasmonic,wu2021deep} of plasmonic nanoantennas. At the same time, a design of nanoparticles is required in many other fields apart from photonics, such as chemistry or biology. For instance, the DL-based inverse design can be implemented to realise specific interactions~\cite{hassan2020artificial}. However, these topics are beyond the scope of the current paper. 

Moving from isolated nanoantennas to their arrays, we wish to mention the results of the DL-based optimization of halide-perovskite thin solar cells combined with a layer of core-shell metallic nanoparticles~\cite{nelson2019using}. This study employs the fact that core-shell nanoparticles may have two plasmon resonances located in different parts of spectra, and the resonances can be tuned by adjusting structural parameters. As a result, the efficiency of such solar cells can be improved by applying the ANN approach, e.g. by studying configurations of core-shell nanoparticles to maximize optical absorption. Similar ideas are applied to the design of metasurfaces with the optimization of structural parameters of single meta-atoms. Thus, it is natural to move from optimizing isolated nanoantennas to the use of DL approaches for advancing the field of metasurfaces.

\section{Transformative metasurfaces}
\label{sec:metasurfaces}

The fundamental concepts and underlying optical physics of metasurfaces have been extensively explored with the field itself now exhibiting a comprehensive framework. Currently, more advanced engineering with optimization tools is required for moving this field to specific practical applications including bending of light, metalenses, and  metaholograms. A rich variety of ML methods and DL approaches has already been explored for metasurfaces aiming to optimize their required functionalities. Below, we present only a few recent examples of such efforts concentrating on particular applications and specific optical devices and functionalities.

% design procedure in general
{\it Inverse design} of metasurfaces results in selecting  particular shapes and materials of an isolated meta-atom as a specific element of metasurface supercell. In this case, representation of parameters plays an important role, often significantly affecting the result of the subsequent training procedure. Similar to the case of nanoantennas, the design procedure here may result in a set of geometric parameters of meta-atoms with fixed shapes. For example, such an approach was used for a metasurface with super-cells consisting of gold cross-shaped resonators (up to 25 resonators per supercell)~\cite{yeung2021multiplexed}. Inputs to the DL algorithm were the corresponding sets of lengths, and the outputs were the absorption spectra. As a result, the algorithm allowed the design of narrow-band, broadband, and multiresonant absorbers.
However, even with adjustment these constrained parameters may fail to provide a desired spectral response. As an alternative, a free-form design approach can be used. In this case, the DL algorithm process is based not on sets of parameters, but rather employs pixelated images of the unit cells~\cite{zhu2021phasetopattern}. Such techniques extend the expressivity of this approach, significantly extending the range of possible cell geometries. Conversely, the DL approach can also be used to restore a unit cell geometry from a given spectra~\cite{malkiel2021inverse}. The design procedure may also incorporate some transfer learning between meta-atoms of various shapes~\cite{xu2021efficient}.

\begin{figure*}[htbp!]
    \centering
    \includegraphics[width=\linewidth]{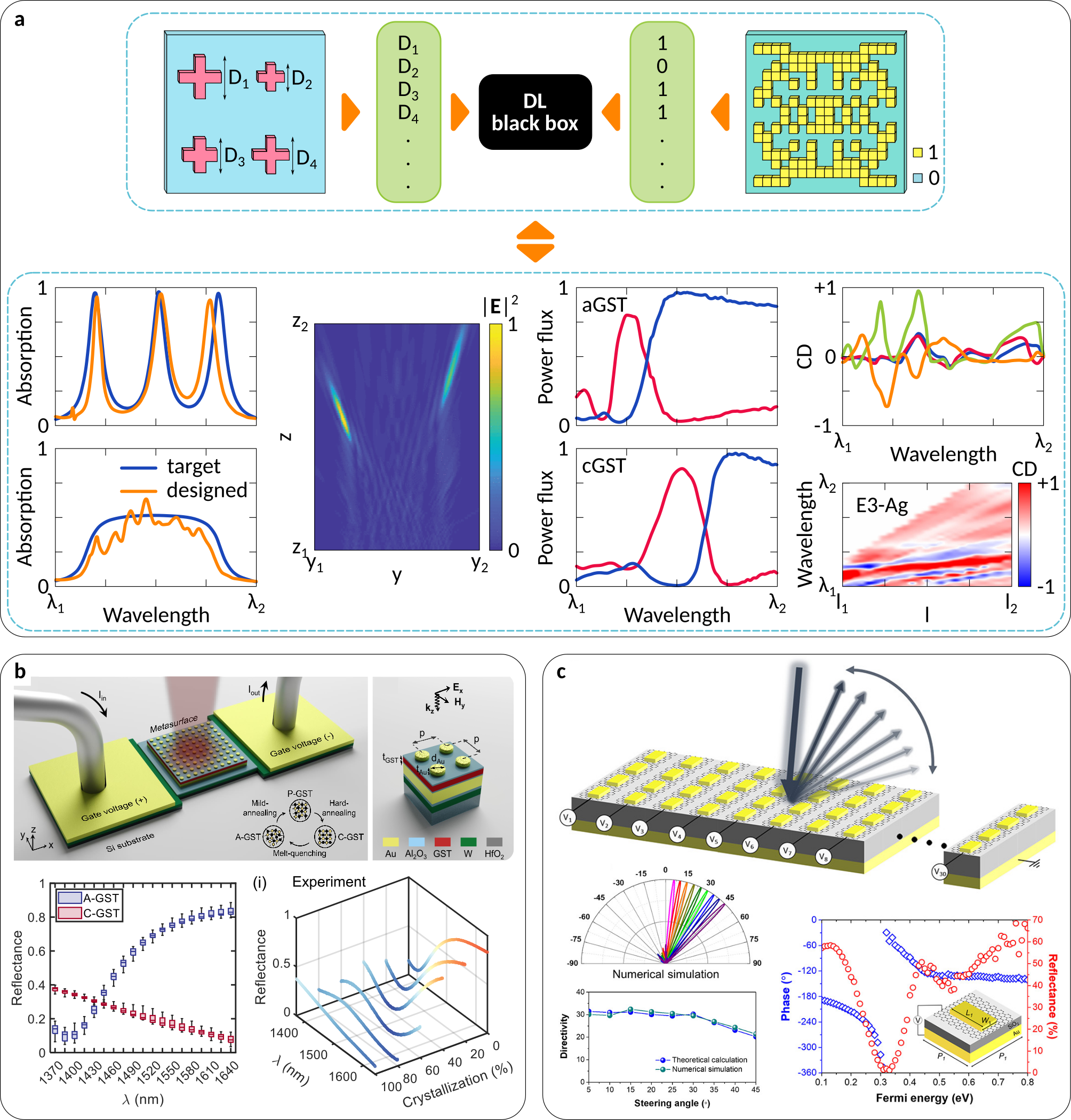}
    \caption{{\bf ML-empowered designs of transformative metasurfaces}. (a) General scheme of the inverse design implemented for metasurfaces. (Adapted from Refs.~\cite{yeung2021multiplexed,zhu2021phasetopattern,gu2021independent,thompson2020artificial,zhang2021graphicprocessable}. Representation of metasurface elements can be done in several ways. For example, DL may process parameters of unit cells with a fixed geometry. At the same time, free-form design can be implemented, such as unit cells are represented as sets of pixiliated images. DL-supplemented design of metasurfaces enable achieving of a great variety of devices, such as multi-resonant and broadband asorbers, metalenses with dual independent focal points, switchable reflectors and sturctures with desired circular dichroism. 
    (b) Example of a designed phase-change metasurface switch (Adapted from~Ref.~\cite{abdollahramezani2021electrically}). (c) Example of a designed beam-steering metasurface (Adapted from Ref.~\cite{lin2021automatic}.}
    \label{fig:metasurfaces}
\end{figure*}

One of the critical tasks in the development of transformative metadevices is to engineer specific properties of the light-matter interaction. As a consequence, many studies are devoted to the design and optimization of specific absorptive, scattering, and diffracting properties~\cite{zhu2021building,zhelyeznyakov2021deep,colburn2021inverse,an2021multifunctional}. However, many studies focus more on design methods rather than  specific applications. We would like to highlight several examples of particular devices benefiting from the AI approach.

Starting from absorbers, there are a variety of realised design procedures resulting in the development of perfect~\cite{han2021metamaterial}, multi-band~\cite{so2021ondemand}, and ultrathin~\cite{chen2021absorption} absorbers. Such studies cover both plasmonic~\cite{sajedian2019finding,lin2020inverse} and dielectric~\cite{deng2021neuraladjoint} metastructures. DL also opens the route for the design of biology-inspired devices such as moth-eye structures~\cite{badloe2020biomimetic} with the designed average absorption reaching $90\%$ in the range from $400$~nm to $1600$~nm.
Scattering properties are a subject of many design procedures~\cite{ghorbani2021deep,ghorbani2021deepa,koziel2021machinelearningpowered,koziel2021design,malkiel2021inverse,zandehshahvar2021manifold}, including those devoted to the development of anisotropic~\cite{wang2020deepa} and bianisotropic~\cite{naseri2021combined} metasurfaces, as well as switchable reflectors~\cite{thompson2020artificial}. DL was also exploited for achieving electromagnetically-induced transparency~\cite{huang2021inverse,yuan2021efficient,zhang2021adaptively}. 
Chiral metasurfaces are also among the typical applications of the DL design procedures~\cite{ma2018deeplearningenabled,li2019selflearning,ashalley2020multitask,tao2020optical,tao2020exploiting,zhang2021graphicprocessable}. Phase-amplitude engineering supported by DL algorithms helps to overcome chromatism~\cite{zhu2020overcome}, achieve tunable beam-steering~\cite{lin2021automatic}, and multiplex an aperture~\cite{zhu2020multiplexing} of  metasurfaces.
AI-assisted design enables development and improvement of  broadband achromatic~\cite{an2021broadband,fan2021timeeffective}, bifocal~\cite{gu2021independent}, thermally tunable~\cite{zarei2021inverse}, and RGB~(red, green, blue)
~\cite{elsawy2021multiobjective} metalenses.

% Structural colors
Light-matter interaction plays an important role in our daily life providing such information about an object as its color. In general, coloration of objects may be achieved by using special pigments characterized by different absorption properties. Another approach is to employ {\it structural colors} generated via engineering of diffraction and reflection properties of an object. Apart from surface decoration, structural colors may be used for digital displays, sensing, storage of information, and many other  applications~\cite{daqiqehrezaei2021nanophotonic}.
Metasurfaces are able to absorb or scatter light selectively with particular wavelengths making them are promising candidates for structural color engineering~\cite{lee2018plasmonic}. Methods to design metasurfaces for this specific purpose also include the application of DL algorithms to plasmonic~\cite{baxter2019plasmonic,roberts2021deep} and, more extensively, for dielectric metastructures~\cite{gao2019bidirectional,hemmatyar2019full,huang2019inverse,kalt2019metamodeling,sajedian2019optimisation,gonzalez-alcalde2020engineering}.

% Solar cells
Metasurfaces may also enhance the performance of {\it solar cells}. To further improve capabilities of metasystems, a novel design of silicon solar cell with an active nanopixel metasurface has been suggested~\cite{shameli2021developing}.  The shape was optimized by using DL to solve the forward design problem, over which an optimisation search is performed using a genetic algorithm to find the maximum short circuit current. As a result, the optimized solar cell exhibited a $~2.5$ times larger short circuit current than in any solar cell without metasurfaces, but with the same amount of crystalline silicon. It was demonstrated that the short circuit current was above $12$mA/cm$^2$ for all other polarizations and angles of incidence. 
Additionally, if the absorptive metasurfaces may be made of various materials, the design procedure can exploit transfer learning methods in order to account for the effect of different materials on the spectra~\cite{noureen2021deepa}.

% LIDAR
One more application is {\it laser imaging detection and ranging (LIDAR)} systems~\cite{mcmanamon2019lidar} extensively used in many areas, such as biology, geology, atmosphere studies, astronomy, and many other fields. With the rise of augmented reality technologies and autonomous systems, such as self-driving vehicles, LIDAR systems entered a new stage of development as a tool for object recognition and 3D modeling of an environment. In Ref.~\cite{lio2021lidar}, the authors suggested a device consisting of a metasurface covered by a layer of liquid crystals with high birefringence. The structure is sandwiched between two cover-slips with a transparent, patterned and conductive layer (ITO), serving as electrodes and connected to an field-programmable gate array (FPGA) control processor. The deflection is performed by the metasurface as well as a change of the surrounding media. Applied external voltage changes the orientation of liquid-crystal molecules resulting in a change of the refractive index enabling beam steering in the desired angle range. The metastructures consist of a patterned MoS$_2$ layer with thickness $30$~nm and refractive index $n = 4.3 + 0.034i$. The ANN approach is employed to generate ensembles of topology-optimized metasurfaces for achieving the highest deflection efficiency for a desired deflection at the telecommunication wavelengths. The demonstrated deflection angles were $45^{\circ}$, $55^{\circ}$, and $65^{\circ}$ with an absolute efficiency of $~0.7$ in all cases..

% Near eye displays
AI methods can also empower technologies such as {\it near-eye displays}, required for virtual and augmented reality systems as well as for vision correction. One of the promising platforms for realizing such displays is based on the application of metasurfaces~\cite{lee2018metasurface,lan2019metasurfaces,long2020colorful,wang2021metalens,bayati2021design} employing their ability to control wavefront, phase, amplitude, and polarization of light. 
Design of metasurfaces for such applications can be realized efficiently with the DL methods. The paper~\cite{chen2021nearly} provides an example of a design of a metasurface acting as a beam deflector. It consists of a stack of $L$-layer gratings made of TiO$_2$ with glass nanoridges where each grating layer is $300$~nm thick. The display system transfers a designated scene toward the human eye via grating-covered waveguide. The system consists of image projectors, in-coupling and out-coupling gratings, and a waveguide. Light propagates along the waveguide towards an eye, and the in-coupling grating is illuminated by single or multiple projectors.
The aim of the DL-based metasurface design is to maximize the deflection efficiency for given operating wavelengths, here chosen to be $720$~nm (red), $540$~nm (green), and $432$~nm (blue). For the cases of integrated and separated RGB projectors, the achieved deflection efficiency was about $91\%$ and $95\%$ respectively, and the overall efficiency is about $80\%$. Also, the authors developed pupil expanded out-coupling grating consisting of three sections. For all of three designated diffraction efficiencies ($33\%$, $50\%$, and $100\%$), the design procedure led to the deflection efficiency above $85\%$. 

% Light sails
A very interesting application of metasurfaces includes the development of the so-called {\it light sails} --- structures for a spacecraft propulsion system relied on a force exerted due to the radiation pressure coming from the Sun or a strong laser beam~\cite{vulpetti2015solar}. Recent studies demonstrated that all-dielectric metasurfaces can maximize the acceleration combined with thermal management requirements and passive stabilization mechanisms~\cite{achouri2019solar,siegel2019selfstabilizing,salary2020photonic}, however, such designs require precise engineering of the structures. Along with some conventional techniques~\cite{jin2020inverse,salary2021inverse}, the DL algorithms were considered as candidates to solve this task~\cite{kudyshev2021optimizing}. Aiming to develop a meta-sail for ultralight spacecraft that can reach Proxima Centauri b in approximately 20 years, the authors of Ref.~\cite{kudyshev2021optimizing} designed free-form silicon metastructures to satisfy optical and weight constraints. They demonstrated that optimization algorithms converged to one-dimensional gratings which allow the acceleration distance $D = 1.9 \times 10^9$~m (distance required to reach target velocity $0.2$ of light speed) and mean reflectivity around $0.81$.

Some other practically important applications include thermal emitters~\cite{kudyshev2020machinelearningassisted}, optical memory elements~\cite{wiecha2019pushing}, smart windows~\cite{balin2019training}, programmable switches~\cite{abdollahramezani2021electrically}, and transmission cloaking devices~\cite{zhen2021realizing}. 
Below, we discuss the application of metasurfaces to biosensing where AI methods and DL tools may provide very beneficial outcomes. 

\section{Chemical and biological sensing}
\label{sec:sensing}

Metasurfaces can be employed as a reliable and robust platform for various chemical and biological sensors~\cite{hassan2021review,tseng2021dielectric,zhang2021metasurfaces}. Being one of the most promising and rapidly developing 
fields of modern photonics, biosensing has also utilised AI aided techniques ~\cite{tittl2019metasurfacebased}. In this case, ML techniques may play two major roles. First, metasurfaces can be employed to optimize sensors~\cite{li2019deep, yan2020design, son2021design, moon2020machine}. On the other hand, metasurfaces empowered by ML techniques provide a powerful tool for classification tasks, which can be exploited to analyze output data of a sensor~\cite{torun2021machine,ren2021midinfrared,john-herpin2021infrared,meng2021plasmonic}.

\begin{figure*}[htbp!]
    \centering
    \includegraphics[width=\linewidth]{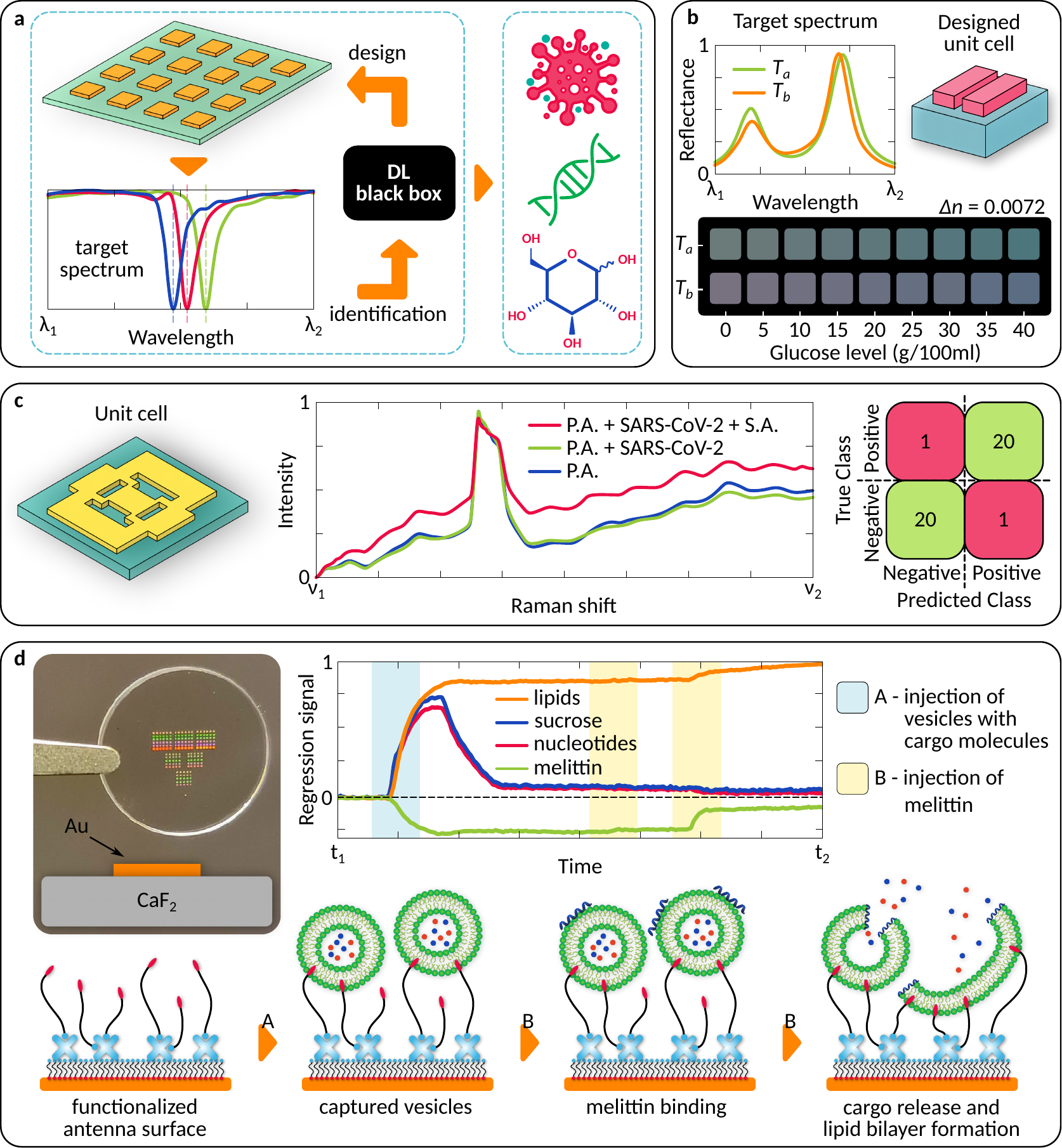}
    \caption{{\bf DL-empowered metasensors.} (a) Schematic of sensing applications of metasurfaces empowered by the of DL methods. DL serves as a tool for designing metasurfaces used as sensing platforms or for analyzing measured response spectra and its classification for the presence of specific molecules, etc. (b) Example of a colorimetric sensor inversely designed with the DL approach (Adapted from Ref.~\cite{son2021design}). Design procedure implies achieving of a target dual-resonant spectra for a double-bar unit cell. As a result, the sensor is capable to distinguish  refractive indexes differed in values by less than $0.01$. 
    (c) DL-supported classification of SARS-CoV-2. Here, unit cell and Raman shift spectra are shown, where P.A. stands for the primary aptamer and S.A. is for the secondary aptamer. Confusion matrix is composed for DL-based classification of clinical samples (Adapted from Ref.~\cite{torun2021machine}).
    (d) Example of a plasmonic sensor for monitoring biomolecule dynamics (Adapted from Ref.~\cite{john-herpin2021infrared}). Real-time analysis of absorbance spectra via ML methods allows to distinguish between dynamically changing biological samples. In particular, regression signal allow tracking dynamics of liposome nanoparticles loaded with sucrose and nucleotides. Introduction of mellittin results in perforation of lipid membrains and cargo release followed by formation of lipid bilayer. This process can be tracked with the help of ML.
    }
    \label{fig:sensors}
\end{figure*}

To demonstrate benefits of the ML-assisted metasensor design, we consider the example of a colorimetric sensor based on all-dielectric metasurfaces. Colorimetry techniques rely on detection of color variations associated with environmental change leading to a change of refractive index. Hence, optimization of colorimetric sensors implies maximization of color difference. In Ref.~\cite{son2021design}, the authors propose a metasurface consisting of double-bar elements. They demonstrated that a dual-resonant type of spectrum can achieve better color difference than a single resonance. Therefore, the design process aims to find target dual-resonance spectra with the highest sensitivity of color change with respect to minute spectral shifts. These spectra were used as inputs to a DL algorithm in order to find geometric parameters of the unit cell. This design procedure is shown in Fig.~\ref{fig:sensors}(a). The optimized structure was used to sense the concentration of a glucose solution. It was demonstrated that the sensing resolution was able to distinguish a change in  refractive index of $~0.0086$ with a mean square error within $0.005$.

Another demonstration of meta-sensor design includes the optimization of double-negative plasmonic metasurfaces for enhanced detection of DNA oligomers~\cite{moon2020machine}. The overall device consisted of a layered structure comprised of a glass substrate, gold layer and  negative-index effective medium. Immobilized and hybridized DNA oligonucleotides were placed on top of the metasurface and covered by a buffer ambience layer (distilled water). A standard multilayer perception was used to predict angular reflectance for a given thickness of the gold layer, effective permittivity and permeability. The obtained spectral characteristics were then clustered to ensure that resonance characteristics are appropriate for an efficient metasensor.
The consequent application of DL and classical ML algorithms resulted in a sensitivity improvement of up to $13$ times, in comparison with conventional plasmonic sensors. 

To illustrate how ML can be employed for classification of biomolecules, we mention the recently suggested platform for surface-enhanced infrared absorption spectroscopy empowered by DL algorithms~\cite{john-herpin2021infrared}. In particular, they use nanoplasmonic metasurface supporting three resonances in a broad mid-IR spectrum ($1000$--$3000$~cm$^{-1}$) covering absorption bands of biomolecules. The sample consists of liposome nanoparticles loaded with sucrose and nucleotides. In addition, melittin is introduced to perforate the lipid membranes of the liposomes leading to dual cargo release and formation of supported lipid bilayer on the surface of the sensor. Real-time measurements of reflectance spectra are used to calculate absorbance which is analyzed via a DL technique.
As a result, the algorithm able to distinguish between each analyte at each point of time [see Fig.~\ref{fig:sensors}(c)]. Therefore, the developed sensing platform enables simultaneous label-free monitoring of major biomolecule classes in water opening the route to study real-time biomolecular interactions, such as vesicle capture, perforation with dual cargo release, and partial transition to planar lipid bilayers.

Advances in metasensors have also been utilised to address global challenges such as COVID-19 pandemic. AI technologies were already exploited for diagnostics of coronavirus~\cite{jamshidi2020artificial,kwekha-rashid2021coronavirus} and recently they were proposed to be combined with meta-sensors. The SARS-CoV-2 saliva sensor [see Fig.~\ref{fig:sensors}(b)] proposed in Ref.\cite{torun2021machine} demonstrated sensitivity and specificity both reaching $95.2\%$ during clinical trials. The sensing platform is comprised of a plasmonic metasurface functionalized with thiol-modified primary DNA aptamers. Meta-atoms were optimized for maximization of the Raman cross-section using a genetic algorithm. Testing samples consisted of unprocessed saliva mixed with Cy5.5-modified fluorescent secondary DNA aptamers. After an allowed 15 minutes for binding, the solution was placed on the sensor to drain. The surface was then rinsed with double-distilled water and phosphate buffer saline. Identification of the virus was performed using an ML algorithm to process Raman shift spectra. Out of $69$ clinical samples only $1$ false positive and $1$ false negative result were obtained indicating a high potential worth further development. Moreover, the sensor was capable of variant detection among inactivated Alpha (B.1.1.7), Beta (B.1.351), and wild-type variants, with cumulative variance $99.7\%$.

We notice that classification of a metasensor output is not necessary done via DL algorithms. Classical ML methods, such as k nearest neighbours or principal component analysis with latent Dirichlet allocation, may provide no less good results. For instance, gaseous and  liquid chemicals may be recognised using plasmonic mid-infrared spectral filters with a photodetector array~\cite{meng2021plasmonic} or plasmonic metasurfaces integrated with microfluidic channel~\cite{ren2021midinfrared}.

We notice also that ML may be used not only to design and supplement metasensors, but also to sense the presence of metamaterials per se. As an example, a DL algorithm could process electromagnetic response signals obtained with THz-band time-domain spectroscopy in order to identify metamaterial in lactose mixtures~\cite{liu2020secure}. The accuracy reached an astonishing value of $100\%$, while classical ML algorithm (support vector machine) allowed only $87.9\%$ and human's ability to recognize the spectrum of a metamaterial was below $57\%$. 

The examples described above, together with other achievements of AI-supplemented sensing, suggest the beginning of a revolutionary era of intelligent biosensors, extensively discussed in literature (see Refs.~\cite{tittl2019metasurfacebased,cui2020advancing,banerjee2021nanostructures,blevins2021roadmap,haick2021artificial} for instance). Seemingly, metastructures and metasurfaces will play an important role in the development of those ideas. Apart from photonics, other sensing technologies have also experienced a rapid onset of AI resulting in the development of next-level intelligent systems based on principles of self-adjustment, which we discuss in the next section.

\section{Self-adapting metasystems}
\label{sec:self_adapting}

Progress in computer science enabled the development of such intriguing concepts as {\it coding} and {\it digital (or programmable) metasurfaces}~\cite{cui2014coding,li2019information,abadal2020programmable,bao2020tunable,cui2020information,ma2020information,tsilipakos2020intelligent,luo2021evolution} that fill a gap between IT and electrodynamics. Though the ML concepts can also be used just for a straightforward design of such structures~\cite{zhang2019machinelearning,banerji2020machinea,shan2020coding,abdullah2021supervisedlearningbased,liu2021intelligent,sui2021deep,yang2021tailoring}, they expose a range of additional possibilities. 
AI technologies which are capable of adjusting the response of a system to specific inputs are extremely useful for the development of {\it self-adapting and re-programmable systems}. In this case, ML can be used as a feedback mechanism, redefining programming sequences of a digital metasurface to adjust to environmental changes.
In recent years, this concept has developed into a new branch of \textit{intelligent metasurfaces}, which rapidly gained a lot of attention for wireless communications at radio frequencies, especially in 6G and internet-of-things technologies~\cite{renzo2019smart,direnzo2020smart,gong2020smart,alexandropoulos2021hybrid,liu2021reconfigurable,long2021promising,munochiveyi2021reconfigurable,wang2021interplay}. 
Below we present only a few examples for illustrating those ideas.

\begin{figure}[htbp!]
  \centering
  \includegraphics[width=0.85\linewidth]{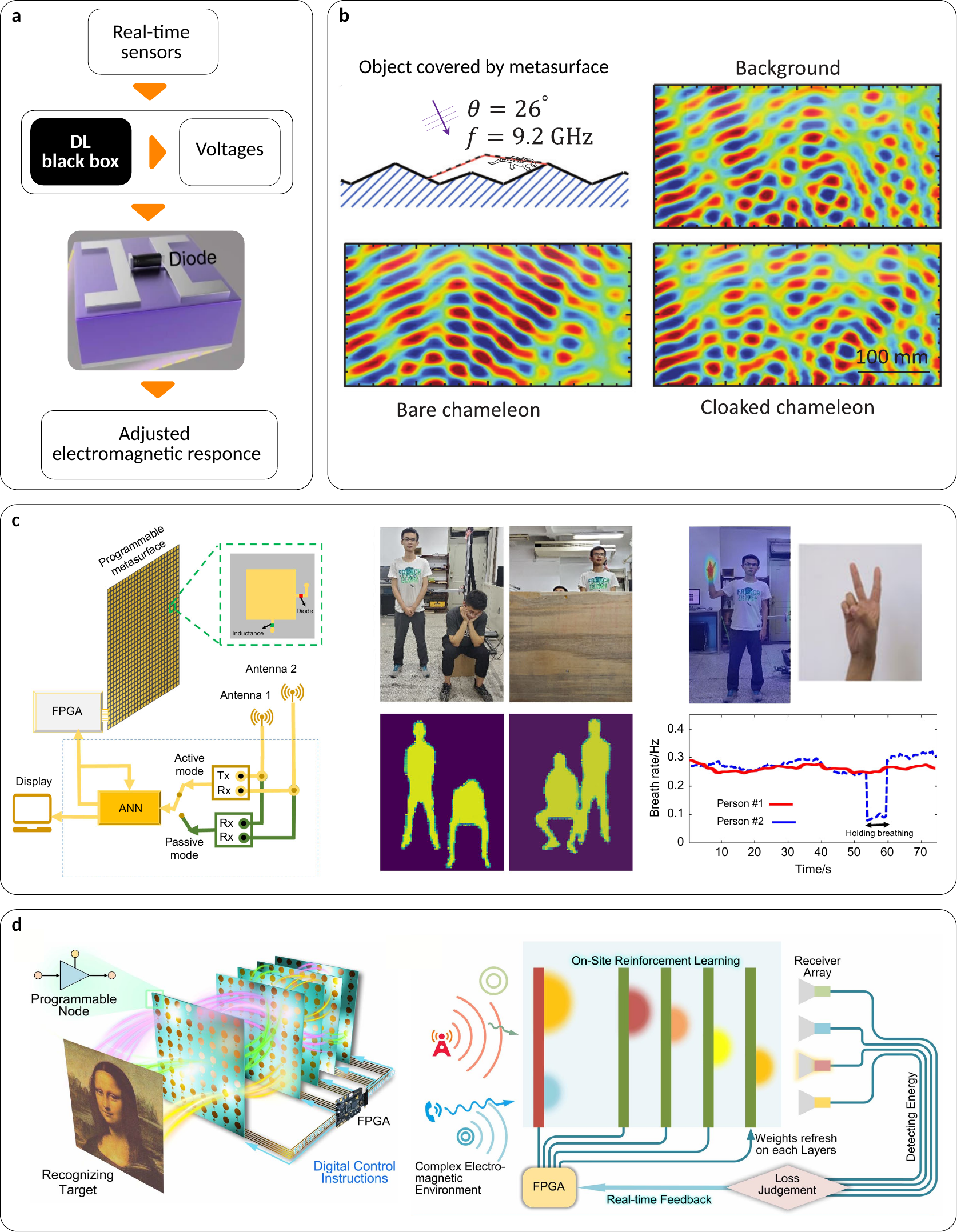}
  \caption{{\bf Self-adapting metasystems and metasurfaces.} (a) General schematic of DL-assisted self-adjusting metadevices. In this case re-programmable metasurfaces are controlled via DL algorithm processing incoming signals. Thus, change of environment leads to change of coding sequence required to adjust the response of a metasurface. 
  (b) Example of a self-adapting metamaterial cloaking device (Adapted from Ref.~\cite{qian2020deeplearningenabled}). An object is covered by a metacloak made of varactor diodes. Sensors perceive changes in background and incoming radiation. This data is processed by DL algorithm determining voltages required to apply to diodes in order to provide cloaking capability.
  (c) DL-assisted microwave imager based on a metasurface (Adapted from Ref. ~\cite{li2019intelligent}). Microwave data coming to a metasurface is processed via DL algorithm to reconstruct image of a human. Another DL algorithm can be used for recognition of specific regions within the reconstructed image, such as hands (gestures). Additionally, the collected data may be used for identification of human breath.
  (d) Metasurface-based optical neural network with re-programmable functions (Adapted from Ref.~\cite{cui2020programmable}). Digital metasurfaces are used as a physical layers of the network. Programming sequences are controlled by FPGA processor. The network is able to dynamically change its functions via re-training procedure. As a result it may be capable of image and feature recognition, coding and de-coding of signals, and other tasks. }
  \label{fig:seld_adapting}
\end{figure}

A self adapting approach was implemented to realize {\it a real-time metasurface imager}~\cite{li2019machinelearning}. The proposed solution is based on $2$-bit coding metasurface --- a structure made of independently controlled meta-atoms supporting four different digital responses, $00$, $01$, $10$ and $11$, which correspond to the physical phases $0$, $\pi/2$, $\pi$ and $3\pi/2$, respectively. Thus, an incident field with different coding sequences will produce different scattering patterns. The principle behind this imaging technique is based on the recognition of an object from the measured scattered fields. In the proposed design, the coding metasurface is controlled via a field programmable gate array. First, the ML algorithms are trained for the desired radiation patterns, and then coding patterns of the metasurface are determined from the obtained radiation patterns.

Another imager was developed in Ref.~\cite{li2019intelligent}. In this case, microwave data collected by a metasurface are used as an input for an ANN reconstructing the image of a human body. A second ANN is used to recognize a particular region within the recorded image, such as a hand or a head. Coding sequences needed for control of  programmable metasurfaces are defined via the Gerchberg-Saxton algorithm. This sequence is implemented to focus radiation waves onto the desired regions of a human body to read the necessary data from the reflected echoes and was demonstrated experimentally. Further development of such systems may lead to next-level human-device interfaces used for smart environments, health monitoring, sign and speech recognition.

{\it Metamaterial cloaks}, or meta-cloaks, previously mentioned in the framework of inverse design, also may be supplemented by ML algorithms enabling self-adapting functions. The cloak presented in Ref.~\cite{qian2020deeplearningenabled} is made of reconfigurable metasurface consisting of varactor diodes. Incoming waves sensed by detectors in real time are processed by a DL algorithm calculating voltages, which have to be applied across the diodes in order to adjust the scattering spectrum. As a result, the engineered scattered field will be similar to that in the case of bare surrounding without an intruding object. The use of DL allows fast determination of the necessary parameters of the metasurface in order to provide cloaking for a wide range of changing parameters, such as the angle and frequency of the incident wave and background variations. As discussed in the literature~\cite{qian2021perspective}, AI technologies may become a key factor in the future development of meta-cloaks and even more fascinating devices. 

Here, we also mention that a relation between AI and photonics is not one-sided. Photonic devices find broad applications for realizing optical ANNs~\cite{shastri2021photonics,xiang2021review} due to their ability to provide parallel computing with the speed of light. Metasurfaces are among a variety of building blocks of such networks~\cite{burgos2021design,luo2021metasurfaceenabled}. Indeed, individual meta-atoms of digital metasurfaces may represent single artificial neurons, controlled via some predefined algorithms. Functionality of optical ANNs may be extended with the use of feedback mechanisms providing re-training of the meta-ANN. For instance, Ref.~\cite{cui2020programmable} demonstrates that coding sequences can be redefined with real-time field-programmable gate arrays. Feedback signals allowed to realize self-learning functions using data from the interaction with environment without prior knowledge. As a result, the authors of Ref.~\cite{cui2020programmable} developed {\it wave-based intelligence machine} able to perform such DL tasks as image recognition and feature detection as well as multi-channel coding and decoding tasks and dynamic multi-beam focusing. We believe that such systems can be developed further by adding a feedback mechanism with another ANN. More discussion on application of metasurfaces for optical neural networks is provided in Sec.~\ref{sec:perspective}.

Moving away from metadevices, it should be mentioned that the AI tools become a part of establishing control systems for lasers~\cite{pu2020automatic}. In this case, ML algorithms are integrated into feedback mechanisms automatically adjusting position and orientation of laser elements, such as waveplates and polarizers. Emergence of similar intelligent systems may change the way how experimental characterization of metasystems and their manufacturing are performed.  Below, 
we discuss briefly some other applications of the AI technologies to other related fields and also provide some perspective on how AI technologies may reshape the development and applications of metadevices.

\section{Perspective and Outlook}
\label{sec:perspective}

Recent advancements in ML and AI methods are expected to reshape some major areas of metaphotonics where metastructures and metasurfaces play important roles. As we mentioned above, to develop metasurfaces with specific properties and functionalities, novel design strategies and approaches in advanced computational techniques are required. It is expected that ML and DL will be useful for developing sophisticated smartphones, robotic systems, and self-driving cars employing the concepts of flat optics. Importantly, ML can help to discover unconventional optical designs thus advancing imaging, sensing, and other functionalities of metaphotonic devices.  Many recent studies are devoted to photonic design approaches and emerging material platforms showcasing ML-assisted optimization for intelligent metasurface designs. 

Among emerging and future developments, we wish to mention a few examples of AI-supplemented systems, beyond the major scope of this paper but still closely related to metaphotonics. 

\begin{figure}
  \centering
  \includegraphics[width=\linewidth]{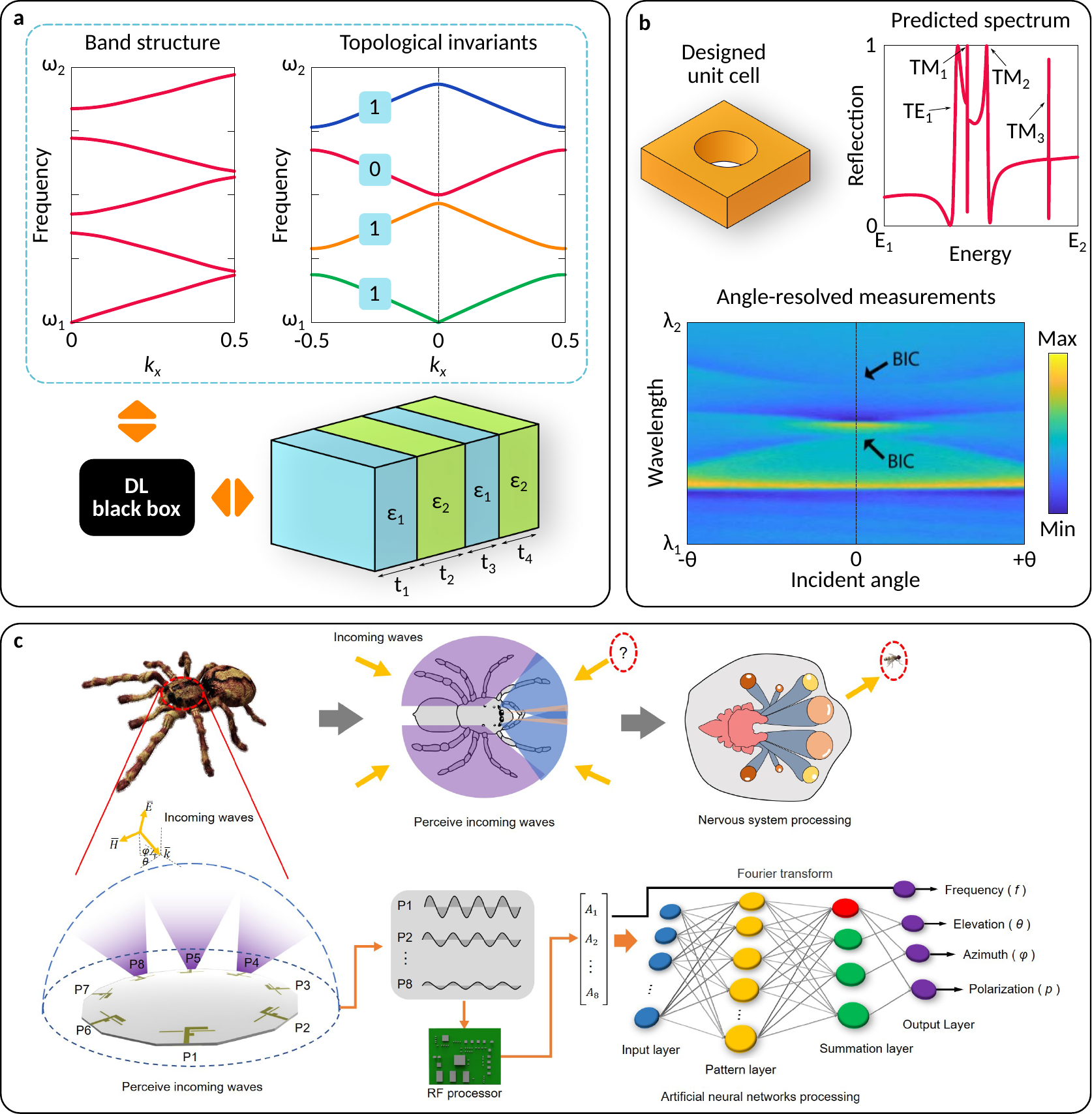}
  \caption{{\bf Examples of DL-empowered systems.} (a) Forward and inverse design procedures for topological properties of 1D photonic crystal. Labels $0$ and $1$ indicate geometric Zak phase of bands (which is either $0$ or $\pi$, correspondingly). (Adopted from Refs.~\cite{singh2020mapping,wu2020machine}).
  (b) Bound states in the continuum designed via DL algorithm, which allowed to predict reflection spectra with automatically labeled modes and find suitable geometric parameters of a unit cell. The results of the design procedure were confirmed experimentally by angle-resolved measurements. (Adopted from Ref.~\cite{ma2021universal}).
  (c) Example of biology-inspired system incorporating DL algorithms. Here, spider-eye-like system is presented, where antenna is used to perceive incoming waves and process them to retrieve information about the environment. ANN in this case explicitly imitates the work of a biological neural network, realizing a system similar to a vision system of a spider. (Adopted from Ref.~\cite{wang2021demonstration}).}
  \label{fig:outlook}
\end{figure}

First, we wish to mention {\it topological photonics}
being one of the most actively developing branches of photonics and physics in general. Realization of topological phases in physical systems opens a road towards novel robust structures protected from scattering losses and structural disorder~\cite{wang2020topological,segev2021topological}.
The current trends of AI-supplemented studies does not leave aside this important area of research.  The first demonstrations of inverse design of photonic topological insulators include the DL-based estimation of geometric parameters of 1D photonic crystal to obtain protected edge states at target frequencies~\cite{pilozzi2018machine,long2019inverse,singh2020mapping}. Later, this approach was also extended to 1D $\mathcal{PT}$-symmetric chains and cylindrical photonic crystal fibers~\cite{pilozzi2021topological}. 
In addition, ANN methods were used to predict topological transitions in photonic crystals~\cite{wu2020machine}. 
Recent works successfully utilizing non-DL methods for inverse design of 2D topological insulators and waveguides~\cite{chen2020inversea,nussbaum2021inverse} indicates that ANNs soon may be applied to design topological structures beyond 1D geometry. At the same time, direction of modern research is turned towards reduction of size and realization of topological metadevices~\cite{zhirihin2021topological} meaning that AI technologies can find their applications in metaphotonics. 

% High-Q resonant states
Metasurfaces supporting {\it high-Q optical resonances} are especially promising for applications in nanophotonics, including enhancement of light-matter interaction, high-harmonic generation, biosensing and nonlinear effects~\cite{koshelev2019metaoptics}. Realization of such opening opportunities require achievement of desired resonance properties. DL-based inverse design of photonic structures supporting BIC was presented in Refs.~\cite{lin2021engineering,ma2021universal}. In particular, in Ref.~\cite{ma2021universal} authors consider suspended photonic crystal slab made of Si$_3$N$_4$ [see Fig.~\ref{fig:outlook}(b)]. Using DL, they find geometric parameters of the structure (radius and height of circular holes) for which symmetry-protected BIC may occur. Importantly, they support the results with experimental measurements demonstrating the presence of two BICs at $\Gamma$-point at wavelengths $700$ and $750$~nm. Widespread attention which BICs receive in the area of photonics~\cite{azzam2021photonic,sadreev2021interference} suggests that more studies supported by AI-algorithms will appear in recent future. 

While the previous sections of this review have discussed the application of DL techniques to photonics, here we briefly discuss an emerging field, namely that of photonic devices as platforms for physical neural networks. Recently there has been increasing interest in approaches termed neuromorphic computing with photonics providing a potentially viable platform \cite{shastri_photonics_2021, schuman_survey_2017}. Neuromorphic computing aims to take an alternative approach to information processing by creating hardware structures that mimic that of nature’s analogue computing i.e. neural structures. This approach is a departure from standard computing models which rely on a centralized processing architecture, instead opting for a distributed model. This approach has several key advantages: firstly this model maps well to information tasks that are distributive, such as neural network models. This match between hardware and algorithm allows an energy efficient approach to information processing. Secondly, for photonics based systems the computation is inherently parallel due to the nature of the platform. Using photonics one can implement different types of neuron schemes with physical implementations having being demonstrated using  Mach–Zehnder interferometers \cite{shen_deep_2017, hughes2018training}, phase change materials \cite{feldmann_all-optical_2019} and diffractive elements \cite{lin_all-optical_2018}. Recently \cite{wright_deep_2021} increased the scope of such applications by proposing a physics aware training scheme that allows a wider class of systems to be trained for DL applications. As metaphotonics develops, it is likely that many of the mechanics and non-linearities possible with metaphotonics will be identified as potential candidates for DL platforms. A metaphotonic hybrid approach to computational sensors has already been demonstrated that operates similar to that of conventional convolutional neural networks \cite{majumdar_metaphotonic_2020}.

Similar to neuromorphic structures there has been development of {\it biology-inspired systems} complimented with AI technologies. As an example, the spider-eye-like antenna demonstrated in Ref.~\cite{wang2021demonstration} is supplemented by a generalised regression neural network \cite{specht_brief_nodate} to analyze the incoming waves (see Fig.~\ref{fig:outlook}(a)). In particular, this algorithm predicts the direction of arrival and the polarization state, simulating a work of a spider's neural system. Similar ideas were implemented for development of AI photonic synapse~\cite{lee2021retinainspired}. The structure is based on a filed-effect transistor with a floating gate of self-assembled perovskite nanocones  embedded into a self-assembled block co-polymer. Arrays of such synapses imitate human retina with position-dependent photosynaptic performance defined by the spatial distribution of the nanocones. 
Such retina represents a single-layer neural network capable of image recognition with up to $90\%$ accuracy. We believe that intelligent metasurfaces may be integrated in such kind of systems to improve their performance and extend functionality, providing additional re-programmable functions or data processing with optical ANNs.

The use of DL methods is not limited to a design of nanoantennas and their optical response. It may become an extremely useful assisting tool for experimental measurements. The demonstrations in this field include characterization of orientation~\cite{hu2020singlenanoparticle} or size~\cite{shiratori2021machinelearned} of metallic nanoparticles using measured spectral data. ML find its applications in a variety of microscopy and imaging  techniques~\cite{pu2021unlabeled,shao2020machine} as well as for tracking~\cite{wang2021deeplearningassisted}, localization~\cite{speiser2021deep} and analysis of single molecules~\cite{zhang2018analyzing}. The proposed algorithms may become a basis for a fast automated retrieval of parameters of nanoantenna samples.

An additional application for DL methods lies within the realm of optimisation, specifically that of high dimensional systems. Global optimisation is a generally challenging endeavour with many problems becoming intractable with a growing number of parameters in the absence of prior knowledge of the parameter landscape. One approach that leverages the power of DL to generalise high dimensional systems is to perform an online optimisation. In this case an ML agent is put in a type of feedback loop with the agent providing predictions as to the best parameters, whilst updating its belief based on feedback from the system. This approach has been used to optimise the laser cooling sequences of cold atom traps with up to 63 individual controllable parameters \cite{tranter_multiparameter_2018, gupta2021machine}. The use of DL allows one to consider optimisation problems with much higher dimensions compared to conventional approaches such as Gaussian process which are generally limited to smaller parameter spaces \cite{wigley_fast_2016, henson_approaching_2018}. Metaphotonics and metasurfaces in particular provide a range if different parameters and exhibit a rich range of phenomena owing to their subwavelength structuring. This provides an opportunity for direct optimisation both in a design setting and real-time capacity, such as that of programmable metasurfaces. Such approaches are indeed flexible as the objective function may be arbitrarily defined to suit any given purpose.

While metaphotonics structures have been intensively studied in the past decade, many challenges still exist for practical and end-specific utilization. Fabrication becomes a tricky issue for many accuracy-required structures, and it also becomes harder to modulate metasurfaces. With emerging materials, new physics, and advanced nanofabrication techniques, those challenges may be bypassed. We anticipate new insights being delivered by merging the concepts from several fields exploring interdisciplinary concepts of metaphotonics and creating new types of hybrid systems governed by the concepts of flat optics and metamaterial-inspired subwavelength engineering. We expect that the future research in intelligent metaphotonics will significantly broaden the horizons of photonics and offer new perspectives for novel applications beyond our current imagination.

\section*{Acknowledgements}

The authors acknowledge useful comments and suggestions received from A. Chukhrov, A. Slobozhanyuk, I. Melchakova, and P. Belov. Y.K. acknowledges a support from the Strategic Fund of the Australian National University, The Australian Research Council (grants DP200101168 and DP210101292), the US Army International Office (grant FA5209-21-P0034), and the Russian Science Foundation (grant 21-72-30018). A.B. acknowledges a support from the Foundation for the Advancement of Theoretical Physics and Mathematics "BASIS". A.T. acknowledges a support from the Australian Research Council Grant No. CE170100012.

\bibliographystyle{unsrt}
\bibliography{bibliography.bib}

\begin{thebibliography}{100}

\bibitem{radovic2018machine}
Alexander Radovic, Mike Williams, David Rousseau, Michael Kagan, Daniele
  Bonacorsi, Alexander Himmel, Adam Aurisano, Kazuhiro Terao, and Taritree
  Wongjirad.
\newblock Machine learning at the energy and intensity frontiers of particle
  physics.
\newblock {\em Nature}, 560(7716):41--48, August 2018.

\bibitem{schmidt2019recent}
Jonathan Schmidt, M{\'a}rio R.~G. Marques, Silvana Botti, and Miguel A.~L.
  Marques.
\newblock Recent advances and applications of machine learning in solid-state
  materials science.
\newblock {\em npj Computational Materials}, 5(1):1--36, August 2019.

\bibitem{brown2020machine}
Keith~A. Brown, Sarah Brittman, Nicol{\`o} Maccaferri, Deep Jariwala, and
  Umberto Celano.
\newblock Machine {{Learning}} in {{Nanoscience}}: {{Big Data}} at {{Small
  Scales}}.
\newblock {\em Nano Letters}, 20(1):2--10, January 2020.

\bibitem{carrasquilla2020machine}
Juan Carrasquilla.
\newblock Machine learning for quantum matter.
\newblock {\em Advances in Physics: X}, 5(1):1797528, January 2020.

\bibitem{bedolla2020machine}
Edwin Bedolla, Luis~Carlos Padierna, and Ram{\'o}n {Casta{\~n}eda-Priego}.
\newblock Machine learning for condensed matter physics.
\newblock {\em Journal of Physics: Condensed Matter}, 33(5):053001, November
  2020.

\bibitem{campbell2021explosion}
Sawyer~D. Campbell, Ronald~P. Jenkins, Philip~J. O'Connor, and Douglas Werner.
\newblock The {{Explosion}} of {{Artificial Intelligence}} in {{Antennas}} and
  {{Propagation}}: {{How Deep Learning Is Advancing Our State}} of the {{Art}}.
\newblock {\em IEEE Antennas and Propagation Magazine}, 63(3):16--27, June
  2021.

\bibitem{koshelev2021dielectric}
Kirill Koshelev and Yuri Kivshar.
\newblock Dielectric {{Resonant Metaphotonics}}.
\newblock {\em ACS Photonics}, 8(1):102--112, January 2021.

\bibitem{zheludev2012metamaterials}
Nikolay~I. Zheludev and Yuri~S. Kivshar.
\newblock From metamaterials to metadevices.
\newblock {\em Nature Materials}, 11(11):917–924, Nov 2012.

\bibitem{kruk2017functional}
Sergey Kruk and Yuri Kivshar.
\newblock Functional {{Meta}}-{{Optics}} and {{Nanophotonics Governed}} by
  {{Mie Resonances}}.
\newblock {\em ACS Photonics}, 4(11):2638--2649, November 2017.

\bibitem{mohri2018foundations}
Mehryar Mohri, Afshin Rostamizadeh, and Ameet Talwalkar.
\newblock {\em Foundations of Machine Learning}.
\newblock MIT Press, 2 edition, Dec 2018.
\newblock Google-Books-ID: dWB9DwAAQBAJ.

\bibitem{mehta2019highbias}
Pankaj Mehta, Marin Bukov, Ching-Hao Wang, Alexandre G.~R. Day, Clint
  Richardson, Charles~K. Fisher, and David~J. Schwab.
\newblock A high-bias, low-variance introduction to {{Machine Learning}} for
  physicists.
\newblock {\em Physics Reports}, 810:1--124, May 2019.

\bibitem{roscher2020explainable}
Ribana Roscher, Bastian Bohn, Marco~F. Duarte, and Jochen Garcke.
\newblock Explainable {{Machine Learning}} for {{Scientific Insights}} and
  {{Discoveries}}.
\newblock {\em IEEE Access}, 8:42200--42216, 2020.

\bibitem{tanaka2021deep}
Akinori Tanaka, Akio Tomiya, and Koji Hashimoto.
\newblock {\em Deep {{Learning}} and {{Physics}}}.
\newblock Mathematical {{Physics Studies}}. {Springer Singapore}, 2021.

\bibitem{miyanawala_efficient_2017}
Tharindu~P. Miyanawala and Rajeev~K. Jaiman.
\newblock An {Efficient} {Deep} {Learning} {Technique} for the
  {Navier}-{Stokes} {Equations}: {Application} to {Unsteady} {Wake} {Flow}
  {Dynamics}.
\newblock October 2017.

\bibitem{lucor_physics-aware_2021}
Didier Lucor, Atul Agrawal, and Anne Sergent.
\newblock Physics-aware deep neural networks for surrogate modeling of
  turbulent natural convection.
\newblock March 2021.

\bibitem{lim2021maxwellnet}
Joowon Lim and Demetri Psaltis.
\newblock {{MaxwellNet}}: {{Physics}}-driven deep neural network training based
  on {{Maxwell}}'s equations.
\newblock {\em arXiv:2107.06164 [physics]}, July 2021.

\bibitem{goodfellow_deep_2016}
Ian Goodfellow, Yoshua Bengio, and Aaron Courville.
\newblock {\em Deep {Learning}}.
\newblock MIT Press, 2016.

\bibitem{hornik1989multilayer}
Kurt Hornik, Maxwell Stinchcombe, and Halbert White.
\newblock Multilayer feedforward networks are universal approximators.
\newblock {\em Neural Networks}, 2(5):359--366, January 1989.

\bibitem{simonyan_very_2015}
Karen Simonyan and Andrew Zisserman.
\newblock Very {Deep} {Convolutional} {Networks} for {Large}-{Scale} {Image}
  {Recognition}.
\newblock {\em arXiv:1409.1556 [cs]}, April 2015.
\newblock arXiv: 1409.1556.

\bibitem{brown_language_2020}
Tom~B. Brown, Benjamin Mann, Nick Ryder, Melanie Subbiah, Jared Kaplan,
  Prafulla Dhariwal, Arvind Neelakantan, Pranav Shyam, Girish Sastry, Amanda
  Askell, Sandhini Agarwal, Ariel Herbert-Voss, Gretchen Krueger, Tom Henighan,
  Rewon Child, Aditya Ramesh, Daniel~M. Ziegler, Jeffrey Wu, Clemens Winter,
  Christopher Hesse, Mark Chen, Eric Sigler, Mateusz Litwin, Scott Gray,
  Benjamin Chess, Jack Clark, Christopher Berner, Sam McCandlish, Alec Radford,
  Ilya Sutskever, and Dario Amodei.
\newblock Language {Models} are {Few}-{Shot} {Learners}.
\newblock {\em arXiv:2005.14165 [cs]}, July 2020.
\newblock arXiv: 2005.14165.

\bibitem{oord_pixel_2016}
Aaron van~den Oord, Nal Kalchbrenner, and Koray Kavukcuoglu.
\newblock Pixel {Recurrent} {Neural} {Networks}.
\newblock {\em arXiv:1601.06759 [cs]}, August 2016.
\newblock arXiv: 1601.06759.

\bibitem{rumelhart_learning_1986}
David~E. Rumelhart, Geoffrey~E. Hinton, and Ronald~J. Williams.
\newblock Learning representations by back-propagating errors.
\newblock {\em Nature}, 323(6088):533--536, October 1986.
\newblock Bandiera\_abtest: a Cg\_type: Nature Research Journals Number: 6088
  Primary\_atype: Research Publisher: Nature Publishing Group.

\bibitem{li2021directional}
Nannan Li, Yunhe Lai, Shiu~Hei Lam, Haoyuan Bai, Lei Shao, and Jianfang Wang.
\newblock Directional {{Control}} of {{Light}} with {{Nanoantennas}}.
\newblock {\em Advanced Optical Materials}, 9(1):2001081, 2021.

\bibitem{rybin2017high}
Mikhail~V. Rybin, Kirill~L. Koshelev, Zarina~F. Sadrieva, Kirill~B. Samusev,
  Andrey~A. Bogdanov, Mikhail~F. Limonov, and Yuri~S. Kivshar.
\newblock High-\${{Q}}\$ {{Supercavity Modes}} in {{Subwavelength Dielectric
  Resonators}}.
\newblock {\em Physical Review Letters}, 119(24):243901, December 2017.

\bibitem{kildishev2013planar}
Alexander~V. Kildishev, Alexandra Boltasseva, and Vladimir~M. Shalaev.
\newblock Planar photonics with metasurfaces.
\newblock {\em Science}, 339(6125):1232009, Mar 2013.

\bibitem{lalanne2017metalenses}
Philippe Lalanne and Pierre Chavel.
\newblock Metalenses at visible wavelengths: past, present, perspectives.
\newblock {\em Laser \& Photonics Reviews}, 11(3):1600295, 2017.

\bibitem{hsiao2017fundamentals}
Hui-Hsin Hsiao, Cheng~Hung Chu, and Din~Ping Tsai.
\newblock Fundamentals and applications of metasurfaces.
\newblock {\em Small Methods}, 1(4):1600064, 2017.

\bibitem{qiu2021quo}
Cheng-Wei Qiu, Tan Zhang, Guangwei Hu, and Yuri Kivshar.
\newblock Quo vadis, metasurfaces?
\newblock {\em Nano Letters}, 21(13):5461–5474, Jul 2021.

\bibitem{chen2021will}
Wei~Ting Chen and Federico Capasso.
\newblock Will flat optics appear in everyday life anytime soon?
\newblock {\em Applied Physics Letters}, 118(10):100503, Mar 2021.

\bibitem{peurifoy2018nanophotonic}
John Peurifoy, Yichen Shen, Li~Jing, Yi~Yang, Fidel {Cano-Renteria}, Brendan~G.
  DeLacy, John~D. Joannopoulos, Max Tegmark, and Marin Solja{\v c}i{\'c}.
\newblock Nanophotonic particle simulation and inverse design using artificial
  neural networks.
\newblock {\em Science Advances}, 4(6):eaar4206, June 2018.

\bibitem{hu2019robust}
Baiqiang Hu, Bei Wu, Dong Tan, Jing Xu, Jing Xu, Yuntian Chen, and Yuntian
  Chen.
\newblock Robust inverse-design of scattering spectrum in core-shell structure
  using modified denoising autoencoder neural network.
\newblock {\em Optics Express}, 27(25):36276--36285, December 2019.

\bibitem{qiu2020inverse}
Cankun Qiu, Zhi Luo, Xia Wu, Huidong Yang, and Bo~Huang.
\newblock Inverse design of multilayer nanoparticles using artificial neural
  networks and genetic algorithm.
\newblock {\em arXiv:2003.08356 [physics, stat]}, March 2020.

\bibitem{so2019simultaneous}
Sunae So, Jungho Mun, and Junsuk Rho.
\newblock Simultaneous {{Inverse Design}} of {{Materials}} and {{Structures}}
  via {{Deep Learning}}: {{Demonstration}} of {{Dipole Resonance Engineering
  Using Core}}\textendash{{Shell Nanoparticles}}.
\newblock {\em ACS Applied Materials \& Interfaces}, 11(27):24264--24268, July
  2019.

\bibitem{qin2019designing}
Feifei Qin, Feifei Qin, Dasen Zhang, Dasen Zhang, Zhenzhen Liu, Zhenzhen Liu,
  Qiang Zhang, Qiang Zhang, Junjun Xiao, and Junjun Xiao.
\newblock Designing metal-dielectric nanoantenna for unidirectional scattering
  via {{Bayesian}} optimization.
\newblock {\em Optics Express}, 27(21):31075--31086, October 2019.

\bibitem{wiecha2020deep}
Peter~R. Wiecha and Otto~L. Muskens.
\newblock Deep {{Learning Meets Nanophotonics}}: {{A Generalized Accurate
  Predictor}} for {{Near Fields}} and {{Far Fields}} of {{Arbitrary 3D
  Nanostructures}}.
\newblock {\em Nano Letters}, 20(1):329--338, January 2020.

\bibitem{vahidzadeh2021artificial}
Ehsan Vahidzadeh and Karthik Shankar.
\newblock Artificial {{Neural Network}}-{{Based Prediction}} of the {{Optical
  Properties}} of {{Spherical Core}}\textendash{{Shell Plasmonic
  Metastructures}}.
\newblock {\em Nanomaterials}, 11(3):633, March 2021.

\bibitem{cao2019hybrid}
Zhaolou Cao, Fenping Cui, Fenglin Xian, Chunjie Zhai, and Shixin Pei.
\newblock A hybrid approach using machine learning and genetic algorithm to
  inverse modeling for single sphere scattering in a {{Gaussian}} light sheet.
\newblock {\em Journal of Quantitative Spectroscopy and Radiative Transfer},
  235:180--186, September 2019.

\bibitem{li2020predicting}
Y.~Li, Y.~Wang, S.~Qi, Q.~Ren, L.~Kang, S.~D. Campbell, P.~L. Werner, and D.~H.
  Werner.
\newblock Predicting {{Scattering From Complex Nano}}-{{Structures}} via {{Deep
  Learning}}.
\newblock {\em IEEE Access}, 8:139983--139993, 2020.

\bibitem{guo2021physics}
Rui Guo, Zhichao Lin, Tao Shan, Xiaoqian Song, Maokun Li, Fan Yang, Shenheng
  Xu, and Aria Abubakar.
\newblock Physics {{Embedded Deep Neural Network}} for {{Solving Full}}-wave
  {{Inverse Scattering Problems}}.
\newblock {\em IEEE Transactions on Antennas and Propagation}, pages 1--1,
  2021.

\bibitem{lin2021lowfrequency}
Zhichao Lin, Rui Guo, Maokun Li, Aria Abubakar, Tao Zhao, Fan Yang, and
  Shenheng Xu.
\newblock Low-{{Frequency Data Prediction With Iterative Learning}} for
  {{Highly Nonlinear Inverse Scattering Problems}}.
\newblock {\em IEEE Transactions on Microwave Theory and Techniques}, pages
  1--1, 2021.

\bibitem{qie2021realtime}
Jinran Qie, Erfan Khoram, Dianjing Liu, Ming Zhou, and Li~Gao.
\newblock Real-time deep learning design tool for far-field radiation profile.
\newblock {\em Photonics Research}, 9(4):B104--B108, April 2021.

\bibitem{sheverdin2020photonic}
Arsen Sheverdin, Francesco Monticone, and Constantinos Valagiannopoulos.
\newblock Photonic {{Inverse Design}} with {{Neural Networks}}: {{The Case}} of
  {{Invisibility}} in the {{Visible}}.
\newblock {\em Physical Review Applied}, 14(2):024054, August 2020.

\bibitem{luo2021deeplearningenabled}
Jie Luo, Xun Li, Xinyuan Zhang, Xinyuan Zhang, Jiajie Guo, Jiajie Guo, Wei Liu,
  Yun Lai, Yaohui Zhan, Yaohui Zhan, Yaohui Zhan, Min Huang, and Min Huang.
\newblock Deep-learning-enabled inverse engineering of multi-wavelength
  invisibility-to-superscattering switching with phase-change materials.
\newblock {\em Optics Express}, 29(7):10527--10537, March 2021.

\bibitem{blanchard-dionne2021successive}
Andre-Pierre {Blanchard-Dionne} and Olivier J.~F. Martin.
\newblock Successive training of a generative adversarial network for the
  design of an optical cloak.
\newblock {\em OSA Continuum}, 4(1):87--95, January 2021.

\bibitem{pan2010survey}
Sinno~Jialin Pan and Qiang Yang.
\newblock A survey on transfer learning.
\newblock {\em IEEE Transactions on Knowledge and Data Engineering},
  22(10):1345–1359, Oct 2010.

\bibitem{zhuang2021comprehensive}
Fuzhen Zhuang, Zhiyuan Qi, Keyu Duan, Dongbo Xi, Yongchun Zhu, Hengshu Zhu, Hui
  Xiong, and Qing He.
\newblock A comprehensive survey on transfer learning.
\newblock {\em Proceedings of the IEEE}, 109(1):43–76, Jan 2021.

\bibitem{qu2019migrating}
Yurui Qu, Li~Jing, Yichen Shen, Min Qiu, and Marin Solja{\v c}i{\'c}.
\newblock Migrating {{Knowledge}} between {{Physical Scenarios Based}} on
  {{Artificial Neural Networks}}.
\newblock {\em ACS Photonics}, 6(5):1168--1174, May 2019.

\bibitem{qiu2021nanophotonic}
Cankun Qiu, Xia Wu, Xia Wu, Zhi Luo, Huidong Yang, Guannan He, Guannan He,
  Bo~Huang, and Bo~Huang.
\newblock Nanophotonic inverse design with deep neural networks based on
  knowledge transfer using imbalanced datasets.
\newblock {\em Optics Express}, 29(18):28406--28415, August 2021.

\bibitem{elzouka2020interpretable}
Mahmoud Elzouka, Charles Yang, Adrian Albert, Sean Lubner, and Ravi~S. Prasher.
\newblock Interpretable inverse design of particle spectral emissivity using
  machine learning.
\newblock {\em arXiv:2002.04223 [physics]}, February 2020.

\bibitem{he2019plasmonic}
Jing He, Chang He, Chao Zheng, Qian Wang, and Jian Ye.
\newblock Plasmonic nanoparticle simulations and inverse design using machine
  learning.
\newblock {\em Nanoscale}, 11(37):17444--17459, September 2019.

\bibitem{wu2021deep}
Qingxin Wu, Qingxin Wu, Xiaozhong Li, Xiaozhong Li, Li~Jiang, Xiao Xu, Dong
  Fang, Jingjing Zhang, Chunyuan Song, Zongfu Yu, Lianhui Wang, Lianhui Wang,
  Li~Gao, and Li~Gao.
\newblock Deep neural network for designing near- and far-field properties in
  plasmonic antennas.
\newblock {\em Optical Materials Express}, 11(7):1907--1917, July 2021.

\bibitem{hassan2020artificial}
Sergio~A. Hassan.
\newblock Artificial neural networks for the inverse design of nanoparticles
  with preferential nano-bio behaviors.
\newblock {\em The Journal of Chemical Physics}, 153(5):054102, August 2020.

\bibitem{nelson2019using}
Michael~D. Nelson and Marcel Di~Vece.
\newblock Using a {{Neural Network}} to {{Improve}} the {{Optical Absorption}}
  in {{Halide Perovskite Layers Containing Core}}-{{Shells Silver
  Nanoparticles}}.
\newblock {\em Nanomaterials}, 9(3):437, March 2019.

\bibitem{yeung2021multiplexed}
Christopher Yeung, Ju-Ming Tsai, Brian King, Benjamin Pham, David Ho, Julia
  Liang, Mark~W. Knight, and Aaswath~P. Raman.
\newblock Multiplexed supercell metasurface design and optimization with tandem
  residual networks.
\newblock {\em Nanophotonics}, 10(3):1133--1143, January 2021.

\bibitem{zhu2021phasetopattern}
Ruichao Zhu, Tianshuo Qiu, Jiafu Wang, Sai Sui, Chenglong Hao, Tonghao Liu,
  Yongfeng Li, Mingde Feng, Anxue Zhang, Cheng-Wei Qiu, and Shaobo Qu.
\newblock Phase-to-pattern inverse design paradigm for fast realization of
  functional metasurfaces via transfer learning.
\newblock {\em Nature Communications}, 12(1):2974, May 2021.

\bibitem{malkiel2021inverse}
Itzik Malkiel, Michael Mrejen, Lior Wolf, and Haim Suchowski.
\newblock Inverse design of unparametrized nanostructures by generating images
  from spectra.
\newblock {\em Optics Letters}, 46(9):2087--2090, May 2021.

\bibitem{xu2021efficient}
Dong Xu, Dong Xu, Yu~Luo, Jun Luo, Jun Luo, Mingbo Pu, Mingbo Pu, Yaxin Zhang,
  Yaxin Zhang, Yinli Ha, Yinli Ha, Xiangang Luo, and Xiangang Luo.
\newblock Efficient design of a dielectric metasurface with transfer learning
  and genetic algorithm.
\newblock {\em Optical Materials Express}, 11(7):1852--1862, July 2021.

\bibitem{gu2021independent}
Yijie Gu, Ran Hao, and Er-Ping Li.
\newblock Independent {{Bifocal Metalens Design Based}} on {{Deep Learning
  Algebra}}.
\newblock {\em IEEE Photonics Technology Letters}, 33(8):403--406, April 2021.

\bibitem{thompson2020artificial}
J.~R. Thompson, J.~R. Thompson, J.~A. Burrow, P.~J. Shah, P.~J. Shah,
  J.~Slagle, E.~S. Harper, A.~Van Rynbach, I.~Agha, and M.~S. Mills.
\newblock Artificial neural network discovery of a switchable metasurface
  reflector.
\newblock {\em Optics Express}, 28(17):24629--24656, August 2020.

\bibitem{zhang2021graphicprocessable}
Jun Zhang, Yukun Luo, Zilong Tao, and Jie You.
\newblock Graphic-processable deep neural network for the efficient prediction
  of {{2D}} diffractive chiral metamaterials.
\newblock {\em Applied Optics}, 60(19):5691--5698, July 2021.

\bibitem{abdollahramezani2021electrically}
Sajjad Abdollahramezani, Omid Hemmatyar, Mohammad Taghinejad, Hossein
  Taghinejad, Alex Krasnok, Ali~A. Eftekhar, Christian Teichrib, Sanchit
  Deshmukh, Mostafa {El-Sayed}, Eric Pop, Matthias Wuttig, Andrea Alu, Wenshan
  Cai, and Ali Adibi.
\newblock Electrically driven programmable phase-change meta-switch reaching
  80\% efficiency.
\newblock {\em arXiv:2104.10381 [physics]}, April 2021.

\bibitem{lin2021automatic}
Chia-Hsiang Lin, Yu-Sheng Chen, Jhao-Ting Lin, Hao~Chung Wu, Hsuan-Ting Kuo,
  Chen-Fu Lin, Peter Chen, and Pin~Chieh Wu.
\newblock Automatic {{Inverse Design}} of {{High}}-{{Performance
  Beam}}-{{Steering Metasurfaces}} via {{Genetic}}-type {{Tree Optimization}}.
\newblock {\em Nano Letters}, 21(12):4981--4989, June 2021.

\bibitem{zhu2021building}
Dayu Zhu, Zhaocheng Liu, Lakshmi Raju, Andrew~S. Kim, Wenshan Cai, and Wenshan
  Cai.
\newblock Building {{Multi}}-functional {{Meta}}-optic {{Systems}} through
  {{Deep Learning}}.
\newblock In {\em Conference on {{Lasers}} and {{Electro}}-{{Optics}} (2021),
  Paper {{FTu4H}}.3}, page FTu4H.3. {Optical Society of America}, May 2021.

\bibitem{zhelyeznyakov2021deep}
Maksym~V. Zhelyeznyakov, Steve Brunton, and Arka Majumdar.
\newblock Deep {{Learning}} to {{Accelerate Scatterer}}-to-{{Field Mapping}}
  for {{Inverse Design}} of {{Dielectric Metasurfaces}}.
\newblock {\em ACS Photonics}, 8(2):481--488, February 2021.

\bibitem{colburn2021inverse}
Shane Colburn and Arka Majumdar.
\newblock Inverse design and flexible parameterization of meta-optics using
  algorithmic differentiation.
\newblock {\em Communications Physics}, 4(1):1--11, March 2021.

\bibitem{an2021multifunctional}
Sensong An, Bowen Zheng, Hong Tang, Mikhail~Y. Shalaginov, Li~Zhou, Hang Li,
  Myungkoo Kang, Kathleen~A. Richardson, Tian Gu, Juejun Hu, Clayton Fowler,
  and Hualiang Zhang.
\newblock Multifunctional {{Metasurface Design}} with a {{Generative
  Adversarial Network}}.
\newblock {\em Advanced Optical Materials}, 9(5):2001433, 2021.

\bibitem{han2021metamaterial}
Cheng Han, Cheng Han, Baifu Zhang, Baifu Zhang, Hao Wang, Jianping Ding, and
  Jianping Ding.
\newblock Metamaterial perfect absorber with morphology-engineered meta-atoms
  using deep learning.
\newblock {\em Optics Express}, 29(13):19955--19963, June 2021.

\bibitem{so2021ondemand}
Sunae So, Younghwan Yang, Taejun Lee, Junsuk Rho, Junsuk Rho, and Junsuk Rho.
\newblock On-demand design of spectrally sensitive multiband absorbers using an
  artificial neural network.
\newblock {\em Photonics Research}, 9(4):B153--B158, April 2021.

\bibitem{chen2021absorption}
Jian Chen, Wei Ding, Ximing Li, Xiang Xi, Kangping Ye, Huabing Wu, and Rui-xin
  Wu.
\newblock Absorption and {{Diffusion Enabled Ultrathin Broadband Metamaterial
  Absorber Designed}} by {{Deep Neural Network}} and {{PSO}}.
\newblock {\em IEEE Antennas and Wireless Propagation Letters}, pages 1--1,
  2021.

\bibitem{sajedian2019finding}
Iman Sajedian, Jeonghyun Kim, and Junsuk Rho.
\newblock Finding the optical properties of plasmonic structures by image
  processing using a combination of convolutional neural networks and recurrent
  neural networks.
\newblock {\em Microsystems \& Nanoengineering}, 5(1):1--8, June 2019.

\bibitem{lin2020inverse}
Ronghui Lin, Yanfen Zhai, Chenxin Xiong, and Xiaohang Li.
\newblock Inverse design of plasmonic metasurfaces by convolutional neural
  network.
\newblock {\em Optics Letters}, 45(6):1362--1365, March 2020.

\bibitem{deng2021neuraladjoint}
Yang Deng, Simiao Ren, Kebin Fan, Jordan~M. Malof, and Willie~J. Padilla.
\newblock Neural-adjoint method for the inverse design of all-dielectric
  metasurfaces.
\newblock {\em Optics Express}, 29(5):7526--7534, March 2021.

\bibitem{badloe2020biomimetic}
Trevon Badloe, Inki Kim, and Junsuk Rho.
\newblock Biomimetic ultra-broadband perfect absorbers optimised with
  reinforcement learning.
\newblock {\em Physical Chemistry Chemical Physics}, 22(4):2337--2342, January
  2020.

\bibitem{ghorbani2021deep}
Fardin Ghorbani, Sina Beyraghi, Javad Shabanpour, Homayoon Oraizi, Hossein
  Soleimani, and Mohammad Soleimani.
\newblock Deep neural network-based automatic metasurface design with a wide
  frequency range.
\newblock {\em Scientific Reports}, 11(1):7102, March 2021.

\bibitem{ghorbani2021deepa}
Fardin Ghorbani, Javad Shabanpour, Sina Beyraghi, Hossein Soleimani, Homayoon
  Oraizi, and Mohammad Soleimani.
\newblock A deep learning approach for inverse design of the metasurface for
  dual-polarized waves.
\newblock {\em arXiv:2105.08508 [physics]}, July 2021.

\bibitem{koziel2021machinelearningpowered}
Slawomir Koziel and Muhammad Abdullah.
\newblock Machine-{{Learning}}-{{Powered EM}}-{{Based Framework}} for
  {{Efficient}} and {{Reliable Design}} of {{Low Scattering Metasurfaces}}.
\newblock {\em IEEE Transactions on Microwave Theory and Techniques},
  69(4):2028--2041, April 2021.

\bibitem{koziel2021design}
Slawomir Koziel, Muhammad Abdullah, and Stanislaw Szczepanski.
\newblock Design of {{High}}-{{Performance Scattering Metasurfaces Through
  Optimization}}-{{Based Explicit RCS Reduction}}.
\newblock {\em IEEE Access}, 9:113077--113088, 2021.

\bibitem{zandehshahvar2021manifold}
Mohammadreza Zandehshahvar, Yashar Kiarashi, Muliang Zhu, Hossein Maleki, Tyler
  Brown, and Ali Adibi.
\newblock Manifold {{Learning}} for {{Knowledge Discovery}} and {{Intelligent
  Inverse Design}} of {{Photonic Nanostructures}}: {{Breaking}} the {{Geometric
  Complexity}}.
\newblock {\em arXiv:2102.04454 [physics]}, February 2021.

\bibitem{wang2020deepa}
Hai~Peng Wang, Yun~Bo Li, He~Li, Shu~Yue Dong, Che Liu, Shi Jin, and Tie~Jun
  Cui.
\newblock Deep {{Learning Designs}} of {{Anisotropic Metasurfaces}} in
  {{Ultrawideband Based}} on {{Generative Adversarial Networks}}.
\newblock {\em Advanced Intelligent Systems}, 2(9):2000068, 2020.

\bibitem{naseri2021combined}
Parinaz Naseri, Stewart Pearson, Zhengzheng Wang, and Sean~V. Hum.
\newblock A {{Combined Machine}}-{{Learning}} / {{Optimization}}-{{Based
  Approach}} for {{Inverse Design}} of {{Nonuniform Bianisotropic
  Metasurfaces}}.
\newblock {\em arXiv:2105.14133 [physics]}, May 2021.

\bibitem{huang2021inverse}
Wei Huang, Ziming Wei, Benying Tan, Shan Yin, and Wentao Zhang.
\newblock Inverse engineering of electromagnetically induced transparency in
  terahertz metamaterial via deep learning.
\newblock {\em Journal of Physics D: Applied Physics}, 54(13):135102, January
  2021.

\bibitem{yuan2021efficient}
Lin Yuan, Lan Wang, Xue-Song Yang, Hao Huang, and Bing-Zhong Wang.
\newblock An {{Efficient Artificial Neural Network Model}} for {{Inverse
  Design}} of {{Metasurfaces}}.
\newblock {\em IEEE Antennas and Wireless Propagation Letters},
  20(6):1013--1017, June 2021.

\bibitem{zhang2021adaptively}
Zhen Zhang, Dai Han, Liuyang Zhang, Xianqiao Wang, and Xuefeng Chen.
\newblock Adaptively reverse design of terahertz metamaterial for
  electromagnetically induced transparency with generative adversarial network.
\newblock {\em Journal of Applied Physics}, 130(3):033101, July 2021.

\bibitem{ma2018deeplearningenabled}
Wei Ma, Feng Cheng, and Yongmin Liu.
\newblock Deep-{{Learning}}-{{Enabled On}}-{{Demand Design}} of {{Chiral
  Metamaterials}}.
\newblock {\em ACS Nano}, 12(6):6326--6334, June 2018.

\bibitem{li2019selflearning}
Yu~Li, Youjun Xu, Meiling Jiang, Bowen Li, Tianyang Han, Cheng Chi, Feng Lin,
  Bo~Shen, Xing Zhu, Luhua Lai, and Zheyu Fang.
\newblock Self-{{Learning Perfect Optical Chirality}} via a {{Deep Neural
  Network}}.
\newblock {\em Physical Review Letters}, 123(21):213902, November 2019.

\bibitem{ashalley2020multitask}
Eric Ashalley, Kingsley Acheampong, Lucas~V. Besteiro, Lucas~V. Besteiro, Peng
  Yu, Arup Neogi, Alexander~O. Govorov, Alexander~O. Govorov, and Zhiming~M.
  Wang.
\newblock Multitask deep-learning-based design of chiral plasmonic
  metamaterials.
\newblock {\em Photonics Research}, 8(7):1213--1225, July 2020.

\bibitem{tao2020optical}
Zilong Tao, Jie You, Jun Zhang, Xin Zheng, Hengzhu Liu, and Tian Jiang.
\newblock Optical circular dichroism engineering in chiral metamaterials
  utilizing a deep learning network.
\newblock {\em Optics Letters}, 45(6):1403--1406, March 2020.

\bibitem{tao2020exploiting}
Zilong Tao, Jun Zhang, Jie You, Hao Hao, Hao Ouyang, Qiuquan Yan, Shiyin Du,
  Zeyu Zhao, Qirui Yang, Xin Zheng, and Tian Jiang.
\newblock Exploiting deep learning network in optical chirality tuning and
  manipulation of diffractive chiral metamaterials.
\newblock {\em Nanophotonics}, 9(9):2945--2956, June 2020.

\bibitem{zhu2020overcome}
Ruichao Zhu, Jiafu Wang, Jiafu Wang, Tianshuo Qiu, Tianshuo Qiu, Sai Sui,
  Yajuan Han, Yuxiang Jia, Yongfeng Li, Mingbao Yan, Yongqiang Pang, Zhuo Xu,
  and Shaobo Qu.
\newblock Overcome chromatism of metasurface via {{Greedy Algorithm}} empowered
  by self-organizing map neural network.
\newblock {\em Optics Express}, 28(24):35724--35733, November 2020.

\bibitem{zhu2020multiplexing}
Ruichao Zhu, Tianshuo Qiu, Jiafu Wang, Sai Sui, Yongfeng Li, Mingde Feng, Hua
  Ma, and Shaobo Qu.
\newblock Multiplexing the aperture of a metasurface: Inverse design via
  deep-learning-forward genetic algorithm.
\newblock {\em Journal of Physics D: Applied Physics}, 53(45):455002, August
  2020.

\bibitem{an2021broadband}
Xipeng An, Yue Cao, Yunxuan Wei, Zhihao Zhou, Tie Hu, Xing Feng, Guangqiang He,
  Ming Zhao, Ming Zhao, and Zhenyu Yang.
\newblock Broadband achromatic metalens design based on deep neural networks.
\newblock {\em Optics Letters}, 46(16):3881--3884, August 2021.

\bibitem{fan2021timeeffective}
Chun-Yuan Fan and Guo-Dung~J. Su.
\newblock Time-{{Effective Simulation Methodology}} for {{Broadband Achromatic
  Metalens Using Deep Neural Networks}}.
\newblock {\em Nanomaterials}, 11(8):1966, August 2021.

\bibitem{zarei2021inverse}
Sanaz Zarei and Amin Khavasi.
\newblock Inverse {{Design}} of {{On}}-{{Chip Thermally Tunable Varifocal
  Metalens Based}} on {{Silicon Metalines}}.
\newblock {\em IEEE Access}, 9:73453--73466, 2021.

\bibitem{elsawy2021multiobjective}
Mahmoud M.~R. Elsawy, Anthony Gourdin, Micka{\"e}l Binois, R{\'e}gis Duvigneau,
  Didier Felbacq, Samira Khadir, Patrice Genevet, and St{\'e}phane Lanteri.
\newblock Multiobjective {{Statistical Learning Optimization}} of {{RGB
  Metalens}}.
\newblock {\em ACS Photonics}, 8(8):2498--2508, August 2021.

\bibitem{daqiqehrezaei2021nanophotonic}
Soroosh Daqiqeh~Rezaei, Zhaogang Dong, John You En~Chan, Jonathan Trisno, Ray
  Jia~Hong Ng, Qifeng Ruan, Cheng-Wei Qiu, N.~Asger Mortensen, and Joel~K.W.
  Yang.
\newblock Nanophotonic {{Structural Colors}}.
\newblock {\em ACS Photonics}, 8(1):18--33, January 2021.

\bibitem{lee2018plasmonic}
Taejun Lee, Jaehyuck Jang, Heonyeong Jeong, and Junsuk Rho.
\newblock Plasmonic- and dielectric-based structural coloring: From
  fundamentals to practical applications.
\newblock {\em Nano Convergence}, 5(1):1, January 2018.

\bibitem{baxter2019plasmonic}
Joshua Baxter, Antonino Cal{\`a}~Lesina, Jean-Michel Guay, Arnaud Weck, Pierre
  Berini, and Lora Ramunno.
\newblock Plasmonic colours predicted by deep learning.
\newblock {\em Scientific Reports}, 9(1):8074, May 2019.

\bibitem{roberts2021deep}
Nathan~Bryn Roberts and Mehdi Keshavarz~Hedayati.
\newblock A deep learning approach to the forward prediction and inverse design
  of plasmonic metasurface structural color.
\newblock {\em Applied Physics Letters}, 119(6):061101, August 2021.

\bibitem{gao2019bidirectional}
Li~Gao, Xiaozhong Li, Dianjing Liu, Lianhui Wang, and Zongfu Yu.
\newblock A {{Bidirectional Deep Neural Network}} for {{Accurate Silicon Color
  Design}}.
\newblock {\em Advanced Materials}, 31(51):1905467, 2019.

\bibitem{hemmatyar2019full}
Omid Hemmatyar, Sajjad Abdollahramezani, Yashar Kiarashinejad, Mohammadreza
  Zandehshahvar, and Ali Adibi.
\newblock Full color generation with {{Fano}}-type resonant {{HfO}} 2
  nanopillars designed by a deep-learning approach.
\newblock {\em Nanoscale}, 11(44):21266--21274, 2019.

\bibitem{huang2019inverse}
Zhao Huang, Xin Liu, and Jianfeng Zang.
\newblock The inverse design of structural color using machine learning.
\newblock {\em Nanoscale}, 11(45):21748--21758, 2019.

\bibitem{kalt2019metamodeling}
Victor Kalt, Alma~K. {Gonz{\'a}lez-Alcalde}, Soukaina {Es-Saidi}, Rafael
  {Salas-Montiel}, Sylvain Blaize, and Demetrio Mac{\'i}as.
\newblock Metamodeling of high-contrast-index gratings for color reproduction.
\newblock {\em JOSA A}, 36(1):79--88, January 2019.

\bibitem{sajedian2019optimisation}
Iman Sajedian, Trevon Badloe, and Junsuk Rho.
\newblock Optimisation of colour generation from dielectric nanostructures
  using reinforcement learning.
\newblock {\em Optics Express}, 27(4):5874--5883, February 2019.

\bibitem{gonzalez-alcalde2020engineering}
Alma~K. {Gonz{\'a}lez-Alcalde}, Rafael {Salas-Montiel}, Victor Kalt, Sylvain
  Blaize, and Demetrio Mac{\'i}as.
\newblock Engineering colors in all-dielectric metasurfaces: Metamodeling
  approach.
\newblock {\em Optics Letters}, 45(1):89--92, January 2020.

\bibitem{shameli2021developing}
Mohammad~Ali Shameli, Amirhossein Fallah, and Leila Yousefi.
\newblock Developing an optimized metasurface for light trapping in thin-film
  solar cells using a deep neural network and a genetic algorithm.
\newblock {\em JOSA B}, 38(9):2728--2735, September 2021.

\bibitem{noureen2021deepa}
Sadia Noureen, Muhammad Zubair, Muhammad Zubair, Mohsen Ali, Muhammad~Qasim
  Mehmood, and Muhammad~Qasim Mehmood.
\newblock Deep learning based hybrid sequence modeling for optical response
  retrieval in metasurfaces for {{STPV}} applications.
\newblock {\em Optical Materials Express}, 11(9):3178--3193, September 2021.

\bibitem{mcmanamon2019lidar}
Paul~F. McManamon.
\newblock {\em {{LiDAR Technologies}} and {{Systems}}}.
\newblock {SPIE Press}, 2019.

\bibitem{lio2021lidar}
Giuseppe~Emanuele Lio and Antonio Ferraro.
\newblock {{LIDAR}} and {{Beam Steering Tailored}} by {{Neuromorphic
  Metasurfaces Dipped}} in a {{Tunable Surrounding Medium}}.
\newblock {\em Photonics}, 8(3):65, March 2021.

\bibitem{lee2018metasurface}
Gun-Yeal Lee, Jong-Young Hong, SoonHyoung Hwang, Seokil Moon, Hyeokjung Kang,
  Sohee Jeon, Hwi Kim, Jun-Ho Jeong, and Byoungho Lee.
\newblock Metasurface eyepiece for augmented reality.
\newblock {\em Nature Communications}, 9(1):4562, November 2018.

\bibitem{lan2019metasurfaces}
Shoufeng Lan, Xueyue Zhang, Mohammad Taghinejad, Sean Rodrigues, Kyu-Tae Lee,
  Zhaocheng Liu, and Wenshan Cai.
\newblock Metasurfaces for {{Near}}-{{Eye Augmented Reality}}.
\newblock {\em ACS Photonics}, 6(4):864--870, April 2019.

\bibitem{long2020colorful}
Shang-Yu Long, Ding-Yue Zhang, Zhen-Zhen Liu, and Jun-Jun Xiao.
\newblock Colorful see-through near-eye display based on all-dielectric
  metasurface.
\newblock In {\em Optics {{Frontier Online}} 2020: {{Optics Imaging}} and
  {{Display}}}, volume 11571, pages 32--39. {SPIE}, October 2020.

\bibitem{wang2021metalens}
Chang Wang, Zeqing Yu, Qiangbo Zhang, Yan Sun, Chenning Tao, Fei Wu, and
  Zhenrong Zheng.
\newblock Metalens {{Eyepiece}} for {{3D Holographic Near}}-{{Eye Display}}.
\newblock {\em Nanomaterials}, 11(8):1920, August 2021.

\bibitem{bayati2021design}
Elyas Bayati, Andrew Wolfram, Shane Colburn, Luocheng Huang, Arka Majumdar, and
  Arka Majumdar.
\newblock Design of achromatic augmented reality visors based on composite
  metasurfaces.
\newblock {\em Applied Optics}, 60(4):844--850, February 2021.

\bibitem{chen2021nearly}
Wen-Qing Chen, Da-Sen Zhang, Shang-Yu Long, Zhen-Zhen Liu, and Jun-Jun Xiao.
\newblock Nearly dispersionless multicolor metasurface beam deflector for near
  eye display designed by a physics-driven deep neural network.
\newblock {\em Applied Optics}, 60(13):3947--3953, May 2021.

\bibitem{vulpetti2015solar}
Giovanni Vulpetti, Les Johnson, and Gregory~L. Matloff.
\newblock {\em Solar {{Sails}}: A {{Novel Approach}} to {{Interplanetary
  Travel}}}.
\newblock Space {{Exploration}}. {Springer-Verlag}, 2 edition.

\bibitem{achouri2019solar}
Karim Achouri, Oscar~V. C{\'e}spedes, and Christophe Caloz.
\newblock Solar ``{{Meta}}-{{Sails}}'' for {{Agile Optical Force Control}}.
\newblock {\em IEEE Transactions on Antennas and Propagation},
  67(11):6924--6934, November 2019.

\bibitem{siegel2019selfstabilizing}
Joel Siegel, Anthony~Y. Wang, Sergey~G. Menabde, Mikhail~A. Kats, Min~Seok
  Jang, and Victor~Watson Brar.
\newblock Self-{{Stabilizing Laser Sails Based}} on {{Optical Metasurfaces}}.
\newblock {\em ACS Photonics}, 6(8):2032--2040, August 2019.

\bibitem{salary2020photonic}
Mohammad~Mahdi Salary and Hossein Mosallaei.
\newblock Photonic {{Metasurfaces}} as {{Relativistic Light Sails}} for
  {{Doppler}}-{{Broadened Stable Beam}}-{{Riding}} and {{Radiative Cooling}}.
\newblock {\em Laser \& Photonics Reviews}, 14(8):1900311, 2020.

\bibitem{jin2020inverse}
Weiliang Jin, Wei Li, Meir Orenstein, and Shanhui Fan.
\newblock Inverse {{Design}} of {{Lightweight Broadband Reflector}} for
  {{Relativistic Lightsail Propulsion}}.
\newblock {\em ACS Photonics}, 7(9):2350--2355, September 2020.

\bibitem{salary2021inverse}
Mohammad~Mahdi Salary and Hossein Mosallaei.
\newblock Inverse {{Design}} of {{Diffractive Relativistic Meta}}-{{Sails}} via
  {{Multi}}-{{Objective Optimization}}.
\newblock {\em Advanced Theory and Simulations}, 4(6):2100047, 2021.

\bibitem{kudyshev2021optimizing}
Zhaxylyk~A. Kudyshev, Alexander~V. Kildishev, Vladimir~M. Shalaev, and
  Alexandra Boltasseva.
\newblock Optimizing {{Startshot}} lightsail design: A generative network-based
  approach.
\newblock {\em arXiv:2108.12999 [physics]}, August 2021.

\bibitem{kudyshev2020machinelearningassisted}
Zhaxylyk~A. Kudyshev, Alexander~V. Kildishev, Vladimir~M. Shalaev, and
  Alexandra Boltasseva.
\newblock Machine-learning-assisted metasurface design for high-efficiency
  thermal emitter optimization.
\newblock {\em Applied Physics Reviews}, 7(2):021407, June 2020.

\bibitem{wiecha2019pushing}
Peter~R. Wiecha, Aur{\'e}lie Lecestre, Nicolas Mallet, and Guilhem Larrieu.
\newblock Pushing the limits of optical information storage using deep
  learning.
\newblock {\em Nature Nanotechnology}, 14(3):237--244, March 2019.

\bibitem{balin2019training}
Igal Balin, Valery Garmider, Yi~Long, and Ibrahim Abdulhalim.
\newblock Training artificial neural network for optimization of nanostructured
  {{VO}}{\textsubscript{2}}-based smart window performance.
\newblock {\em Optics Express}, 27(16):A1030--A1040, August 2019.

\bibitem{zhen2021realizing}
Zheng Zhen, Zheng Zhen, Zheng Zhen, Chao Qian, Chao Qian, Chao Qian, Chao Qian,
  Yuetian Jia, Yuetian Jia, Yuetian Jia, Zhixiang Fan, Zhixiang Fan, Zhixiang
  Fan, Ran Hao, Ran Hao, Tong Cai, Tong Cai, Tong Cai, Bin Zheng, Bin Zheng,
  Bin Zheng, Bin Zheng, Hongsheng Chen, Hongsheng Chen, Hongsheng Chen, Erping
  Li, Erping Li, and Erping Li.
\newblock Realizing transmitted metasurface cloak by a tandem neural network.
\newblock {\em Photonics Research}, 9(5):B229--B235, May 2021.

\bibitem{hassan2021review}
Mohammad~Muntasir Hassan, Farhan~Sadik Sium, Fariba Islam, and Sajid~Muhaimin
  Choudhury.
\newblock A review on plasmonic and metamaterial based biosensing platforms for
  virus detection.
\newblock {\em Sensing and Bio-Sensing Research}, 33:100429, August 2021.

\bibitem{tseng2021dielectric}
Ming~Lun Tseng, Yasaman Jahani, Aleksandrs Leitis, and Hatice Altug.
\newblock Dielectric {{Metasurfaces Enabling Advanced Optical Biosensors}}.
\newblock {\em ACS Photonics}, 8(1):47--60, January 2021.

\bibitem{zhang2021metasurfaces}
Shuyan Zhang, Chi~Lok Wong, Shuwen Zeng, Renzhe Bi, Kolvyn Tai, Kishan
  Dholakia, and Malini Olivo.
\newblock Metasurfaces for biomedical applications: Imaging and sensing from a
  nanophotonics perspective.
\newblock {\em Nanophotonics}, 10(1):259--293, January 2021.

\bibitem{tittl2019metasurfacebased}
Andreas Tittl, Aurelian {John-Herpin}, Aleksandrs Leitis, Eduardo~R. Arvelo,
  and Hatice Altug.
\newblock Metasurface-{{Based Molecular Biosensing Aided}} by {{Artificial
  Intelligence}}.
\newblock {\em Angewandte Chemie International Edition}, 58(42):14810--14822,
  2019.

\bibitem{li2019deep}
Xiaozhong Li, Jing Shu, Wenhua Gu, and Li~Gao.
\newblock Deep neural network for plasmonic sensor modeling.
\newblock {\em Optical Materials Express}, 9(9):3857--3862, September 2019.

\bibitem{yan2020design}
Ruoqin Yan, Tao Wang, Xiaoyun Jiang, Qingfang Zhong, Xing Huang, Lu~Wang, and
  Xinzhao Yue.
\newblock Design of high-performance plasmonic nanosensors by particle swarm
  optimization algorithm combined with machine learning.
\newblock {\em Nanotechnology}, 31(37):375202, June 2020.

\bibitem{son2021design}
Hyunwoo Son, Sun-Je Kim, Jongwoo Hong, Jangwoon Sung, and Byoungho Lee.
\newblock Design of {{Highly Perceptible Dual}}-resonance {{All}}-dielectric
  {{Metasurface Colorimetric Sensor}} via {{Deep Neural Networks}}.
\newblock August 2021.

\bibitem{moon2020machine}
Gwiyeong Moon, Jong-ryul Choi, Changhun Lee, Youngjin Oh, Kyung~Hwan Kim, and
  Donghyun Kim.
\newblock Machine learning-based design of meta-plasmonic biosensors with
  negative index metamaterials.
\newblock {\em Biosensors and Bioelectronics}, 164:112335, September 2020.

\bibitem{torun2021machine}
Hulya Torun, Buse Bilgin, Muslum Ilgu, Cenk Yanik, Numan Batur, Suleyman Celik,
  Meric Ozturk, Ozlem Dogan, Onder Ergonul, Ihsan Solaroglu, Fusun Can, and
  Mehmet~Cengiz Onbasli.
\newblock Machine learning detects {{SARS}}-{{CoV}}-2 and variants rapidly on
  {{DNA}} aptamer metasurfaces.
\newblock page 2021.08.07.21261749, August 2021.

\bibitem{ren2021midinfrared}
Zhihao Ren, Zixuan Zhang, Jingxuan Wei, Bowei Dong, and Chengkuo Lee.
\newblock Mid-infrared {{Nanoantennas}} as {{Ultrasensitive Vibrational Probes
  Assisted}} by {{Machine Learning}} and {{Hyperspectral Imaging}}.
\newblock August 2021.

\bibitem{john-herpin2021infrared}
Aurelian {John-Herpin}, Deepthy Kavungal, Lea von M{\"u}cke, and Hatice Altug.
\newblock Infrared {{Metasurface Augmented}} by {{Deep Learning}} for
  {{Monitoring Dynamics}} between {{All Major Classes}} of {{Biomolecules}}.
\newblock {\em Advanced Materials}, 33(14):2006054, 2021.

\bibitem{meng2021plasmonic}
Jiajun Meng, Jasper~J. Cadusch, and Kenneth~B. Crozier.
\newblock Plasmonic {{Mid}}-{{Infrared Filter Array}}-{{Detector Array Chemical
  Classifier Based}} on {{Machine Learning}}.
\newblock {\em ACS Photonics}, 8(2):648--657, February 2021.

\bibitem{jamshidi2020artificial}
Mohammad Jamshidi, Ali Lalbakhsh, Jakub Talla, Zden{\v e}k Peroutka, Farimah
  Hadjilooei, Pedram Lalbakhsh, Morteza Jamshidi, Luigi~La Spada, Mirhamed
  Mirmozafari, Mojgan Dehghani, Asal Sabet, Saeed Roshani, Sobhan Roshani, Nima
  {Bayat-Makou}, Bahare Mohamadzade, Zahra Malek, Alireza Jamshidi, Sarah
  Kiani, Hamed {Hashemi-Dezaki}, and Wahab Mohyuddin.
\newblock Artificial {{Intelligence}} and {{COVID}}-19: {{Deep Learning
  Approaches}} for {{Diagnosis}} and {{Treatment}}.
\newblock {\em IEEE Access}, 8:109581--109595, 2020.

\bibitem{kwekha-rashid2021coronavirus}
Ameer~Sardar {Kwekha-Rashid}, Heamn~N. Abduljabbar, and Bilal Alhayani.
\newblock Coronavirus disease ({{COVID}}-19) cases analysis using
  machine-learning applications.
\newblock {\em Applied Nanoscience}, May 2021.

\bibitem{liu2020secure}
Feifei Liu, Weihao Zhang, Yu~Sun, Jianwei Liu, Jungang Miao, Feng He, and
  Xiaojun Wu.
\newblock Secure {{Deep Learning}} for {{Intelligent Terahertz Metamaterial
  Identification}}.
\newblock {\em Sensors}, 20(19):5673, January 2020.

\bibitem{cui2020advancing}
Feiyun Cui, Yun Yue, Yi~Zhang, Ziming Zhang, and H.~Susan Zhou.
\newblock Advancing {{Biosensors}} with {{Machine Learning}}.
\newblock {\em ACS Sensors}, 5(11):3346--3364, November 2020.

\bibitem{banerjee2021nanostructures}
Aishwaryadev Banerjee, Swagata Maity, and Carlos~H. Mastrangelo.
\newblock Nanostructures for {{Biosensing}}, with a {{Brief Overview}} on
  {{Cancer Detection}}, {{IoT}}, and the {{Role}} of {{Machine Learning}} in
  {{Smart Biosensors}}.
\newblock {\em Sensors}, 21(4):1253, January 2021.

\bibitem{blevins2021roadmap}
Morgan~G. Blevins, Alvaro {Fernandez-Galiana}, Milo~J. Hooper, and Svetlana~V.
  Boriskina.
\newblock Roadmap on {{Universal Photonic Biosensors}} for {{Real}}-{{Time
  Detection}} of {{Emerging Pathogens}}.
\newblock {\em Photonics}, 8(8):342, August 2021.

\bibitem{haick2021artificial}
Hossam Haick and Ning Tang.
\newblock Artificial {{Intelligence}} in {{Medical Sensors}} for {{Clinical
  Decisions}}.
\newblock {\em ACS Nano}, 15(3):3557--3567, March 2021.

\bibitem{cui2014coding}
Tie~Jun Cui, Mei~Qing Qi, Xiang Wan, Jie Zhao, and Qiang Cheng.
\newblock Coding metamaterials, digital metamaterials and programmable
  metamaterials.
\newblock {\em Light: Science \& Applications}, 3(10):e218--e218, October 2014.

\bibitem{li2019information}
Lianlin Li and Tie~Jun Cui.
\newblock Information metamaterials \textendash{} from effective media to
  real-time information processing systems.
\newblock {\em Nanophotonics}, 8(5):703--724, May 2019.

\bibitem{abadal2020programmable}
Sergi Abadal, Tie-Jun Cui, Tony Low, and Julius Georgiou.
\newblock Programmable {{Metamaterials}} for {{Software}}-{{Defined
  Electromagnetic Control}}: {{Circuits}}, {{Systems}}, and {{Architectures}}.
\newblock {\em IEEE Journal on Emerging and Selected Topics in Circuits and
  Systems}, 10(1):6--19, March 2020.

\bibitem{bao2020tunable}
Lei Bao and Tie~Jun Cui.
\newblock Tunable, reconfigurable, and programmable metamaterials.
\newblock {\em Microwave and Optical Technology Letters}, 62(1):9--32, 2020.

\bibitem{cui2020information}
Tie~Jun Cui, Lianlin Li, Shuo Liu, Qian Ma, Lei Zhang, Xiang Wan, Wei~Xiang
  Jiang, and Qiang Cheng.
\newblock Information {{Metamaterial Systems}}.
\newblock {\em iScience}, 23(8):101403, July 2020.

\bibitem{ma2020information}
Qian Ma and Tie~Jun Cui.
\newblock Information {{Metamaterials}}: Bridging the physical world and
  digital world.
\newblock {\em PhotoniX}, 1(1):1, March 2020.

\bibitem{tsilipakos2020intelligent}
Odysseas Tsilipakos, Anna~C. Tasolamprou, Alexandros Pitilakis, Fu~Liu, Xuchen
  Wang, Mohammad~Sajjad Mirmoosa, Dimitrios~C. Tzarouchis, Sergi Abadal,
  Hamidreza Taghvaee, Christos Liaskos, Ageliki Tsioliaridou, Julius Georgiou,
  Albert {Cabellos-Aparicio}, Eduard Alarc{\'o}n, Sotiris Ioannidis, Andreas
  Pitsillides, Ian~F. Akyildiz, Nikolaos~V. Kantartzis, Eleftherios~N.
  Economou, Costas~M. Soukoulis, Maria Kafesaki, and Sergei Tretyakov.
\newblock Toward {{Intelligent Metasurfaces}}: {{The Progress}} from {{Globally
  Tunable Metasurfaces}} to {{Software}}-{{Defined Metasurfaces}} with an
  {{Embedded Network}} of {{Controllers}}.
\newblock {\em Advanced Optical Materials}, 8(17):2000783, 2020.

\bibitem{luo2021evolution}
Sisi Luo, Jianjiao Hao, Fuju Ye, Jiaxin Li, Ying Ruan, Haoyang Cui, Wenjun Liu,
  and Lei Chen.
\newblock Evolution of the {{Electromagnetic Manipulation}}: {{From Tunable}}
  to {{Programmable}} and {{Intelligent Metasurfaces}}.
\newblock {\em Micromachines}, 12(8):988, August 2021.

\bibitem{zhang2019machinelearning}
Qian Zhang, Che Liu, Xiang Wan, Lei Zhang, Shuo Liu, Yan Yang, and Tie~Jun Cui.
\newblock Machine-{{Learning Designs}} of {{Anisotropic Digital Coding
  Metasurfaces}}.
\newblock {\em Advanced Theory and Simulations}, 2(2):1800132, 2019.

\bibitem{banerji2020machinea}
Sourangsu Banerji, Apratim Majumder, Alex Hamrick, Rajesh Menon, and Berardi
  {Sensale-Rodriguez}.
\newblock Machine {{Learning}} enables {{Ultra}}-{{Compact Integrated
  Photonics}} through {{Silicon}}-{{Nanopattern Digital Metamaterials}}.
\newblock {\em arXiv:2011.11754 [physics]}, November 2020.

\bibitem{shan2020coding}
Tao Shan, Xiaotian Pan, Maokun Li, Shenheng Xu, and Fan Yang.
\newblock Coding {{Programmable Metasurfaces Based}} on {{Deep Learning
  Techniques}}.
\newblock {\em IEEE Journal on Emerging and Selected Topics in Circuits and
  Systems}, 10(1):114--125, March 2020.

\bibitem{abdullah2021supervisedlearningbased}
Muhammad Abdullah and Slawomir Koziel.
\newblock Supervised-{{Learning}}-{{Based Development}} of {{Multibit
  RCS}}-{{Reduced Coding Metasurfaces}}.
\newblock {\em IEEE Transactions on Microwave Theory and Techniques}, pages
  1--1, 2021.

\bibitem{liu2021intelligent}
Che Liu, Che Liu, Wen~Ming Yu, Wen~Ming Yu, Qian Ma, Qian Ma, Lianlin Li,
  Tie~Jun Cui, and Tie~Jun Cui.
\newblock Intelligent coding metasurface holograms by physics-assisted
  unsupervised generative adversarial network.
\newblock {\em Photonics Research}, 9(4):B159--B167, April 2021.

\bibitem{sui2021deep}
Fanping Sui, Ruiqi Guo, Zhizhou Zhang, Grace~X. Gu, and Liwei Lin.
\newblock Deep {{Reinforcement Learning}} for {{Digital Materials Design}}.
\newblock {\em ACS Materials Letters}, pages 1433--1439, August 2021.

\bibitem{yang2021tailoring}
Shuai Yang, Kuang Zhang, Xumin Ding, Guohui Yang, and Qun Wu.
\newblock Tailoring the scattering properties of coding metamaterials based on
  machine learning.
\newblock {\em EPJ Applied Metamaterials}, 8:15, 2021.

\bibitem{renzo2019smart}
Marco~Di Renzo, Merouane Debbah, Dinh-Thuy {Phan-Huy}, Alessio Zappone,
  Mohamed-Slim Alouini, Chau Yuen, Vincenzo Sciancalepore, George~C.
  Alexandropoulos, Jakob Hoydis, Haris Gacanin, Julien de~Rosny, Ahcene
  Bounceur, Geoffroy Lerosey, and Mathias Fink.
\newblock Smart radio environments empowered by reconfigurable {{AI}}
  meta-surfaces: An idea whose time has come.
\newblock {\em EURASIP Journal on Wireless Communications and Networking},
  2019(1):129, May 2019.

\bibitem{direnzo2020smart}
Marco Di~Renzo, Alessio Zappone, Merouane Debbah, Mohamed-Slim Alouini, Chau
  Yuen, Julien {de Rosny}, and Sergei Tretyakov.
\newblock Smart {{Radio Environments Empowered}} by {{Reconfigurable
  Intelligent Surfaces}}: {{How It Works}}, {{State}} of {{Research}}, and
  {{The Road Ahead}}.
\newblock {\em IEEE Journal on Selected Areas in Communications},
  38(11):2450--2525, November 2020.

\bibitem{gong2020smart}
Shimin Gong, Xiao Lu, Dinh~Thai Hoang, Dusit Niyato, Lei Shu, Dong~In Kim, and
  Ying-Chang Liang.
\newblock Toward {{Smart Wireless Communications}} via {{Intelligent Reflecting
  Surfaces}}: {{A Contemporary Survey}}.
\newblock {\em IEEE Communications Surveys Tutorials}, 22(4):2283--2314, 2020.

\bibitem{alexandropoulos2021hybrid}
George~C. Alexandropoulos, Nir Shlezinger, Idban Alamzadeh, Mohammadreza~F.
  Imani, Haiyang Zhang, and Yonina~C. Eldar.
\newblock Hybrid {{Reconfigurable Intelligent Metasurfaces}}: {{Enabling
  Simultaneous Tunable Reflections}} and {{Sensing}} for {{6G Wireless
  Communications}}.
\newblock {\em arXiv:2104.04690 [cs, eess, math]}, April 2021.

\bibitem{liu2021reconfigurable}
Yuanwei Liu, Xiao Liu, Xidong Mu, Tianwei Hou, Jiaqi Xu, Marco Di~Renzo, and
  Naofal {Al-Dhahir}.
\newblock Reconfigurable {{Intelligent Surfaces}}: {{Principles}} and
  {{Opportunities}}.
\newblock {\em IEEE Communications Surveys Tutorials}, 23(3):1546--1577, 2021.

\bibitem{long2021promising}
Wenxuan Long, Rui Chen, Marco Moretti, Wei Zhang, and Jiandong Li.
\newblock A {{Promising Technology}} for {{6G Wireless Networks}}:
  {{Intelligent Reflecting Surface}}.
\newblock {\em Journal of Communications and Information Networks}, 6(1):1--16,
  March 2021.

\bibitem{munochiveyi2021reconfigurable}
Munyaradzi Munochiveyi, Arjun~Chakravarthi Pogaku, Dinh-Thuan Do, Anh-Tu Le,
  Miroslav Voznak, and Nhan~Duc Nguyen.
\newblock Reconfigurable {{Intelligent Surface Aided Multi}}-{{User
  Communications}}: {{State}}-of-the-{{Art Techniques}} and {{Open Issues}}.
\newblock {\em IEEE Access}, 9:118584--118605, 2021.

\bibitem{wang2021interplay}
Jinghe Wang, Wankai Tang, Yu~Han, Shi Jin, Xiao Li, Chao-Kai Wen, Qiang Cheng,
  and Tie~Jun Cui.
\newblock Interplay {{Between RIS}} and {{AI}} in {{Wireless Communications}}:
  {{Fundamentals}}, {{Architectures}}, {{Applications}}, and {{Open Research
  Problems}}.
\newblock {\em IEEE Journal on Selected Areas in Communications},
  39(8):2271--2288, August 2021.

\bibitem{qian2020deeplearningenabled}
Chao Qian, Bin Zheng, Yichen Shen, Li~Jing, Erping Li, Lian Shen, and Hongsheng
  Chen.
\newblock Deep-learning-enabled self-adaptive microwave cloak without human
  intervention.
\newblock {\em Nature Photonics}, 14(6):383--390, June 2020.

\bibitem{li2019intelligent}
Lianlin Li, Ya~Shuang, Qian Ma, Haoyang Li, Hanting Zhao, Menglin Wei, Che Liu,
  Chenglong Hao, Cheng-Wei Qiu, and Tie~Jun Cui.
\newblock Intelligent metasurface imager and recognizer.
\newblock {\em Light: Science \& Applications}, 8(1):97, October 2019.

\bibitem{cui2020programmable}
Tie~Jun Cui, Che Liu, qian {ma}, Zhangjie Luo, Qiaoru Hong, Qiang Xiao, Hao~Chi
  Zhang, Long Miao, Wenming Yu, Qiang Cheng, and Lianlin Li.
\newblock Programmable artificial intelligence machine for wave sensing and
  communications.
\newblock October 2020.

\bibitem{li2019machinelearning}
Lianlin Li, Hengxin Ruan, Che Liu, Ying Li, Ya~Shuang, Andrea Al{\`u},
  Cheng-Wei Qiu, and Tie~Jun Cui.
\newblock Machine-learning reprogrammable metasurface imager.
\newblock {\em Nature Communications}, 10(1):1082, March 2019.

\bibitem{qian2021perspective}
Chao Qian and Hongsheng Chen.
\newblock A perspective on the next generation of invisibility
  cloaks\textemdash{{Intelligent}} cloaks.
\newblock {\em Applied Physics Letters}, 118(18):180501, May 2021.

\bibitem{shastri2021photonics}
Bhavin~J. Shastri, Alexander~N. Tait, T.~{Ferreira de Lima}, Wolfram H.~P.
  Pernice, Harish Bhaskaran, C.~D. Wright, and Paul~R. Prucnal.
\newblock Photonics for artificial intelligence and neuromorphic computing.
\newblock {\em Nature Photonics}, 15(2):102--114, February 2021.

\bibitem{xiang2021review}
Shuiying Xiang, Yanan Han, Ziwei Song, Xingxing Guo, Yahui Zhang, Zhenxing Ren,
  Suhong Wang, Yuanting Ma, Weiwen Zou, Bowen Ma, Shaofu Xu, Jianji Dong,
  Hailong Zhou, Quansheng Ren, Tao Deng, Yan Liu, Genquan Han, and Yue Hao.
\newblock A review: {{Photonics}} devices, architectures, and algorithms for
  optical neural computing.
\newblock {\em Journal of Semiconductors}, 42(2):023105, February 2021.

\bibitem{burgos2021design}
Carlos Mauricio~Villegas Burgos, Tianqi Yang, Yuhao Zhu, Yuhao Zhu, and
  A.~Nickolas Vamivakas.
\newblock Design framework for metasurface optics-based convolutional neural
  networks.
\newblock {\em Applied Optics}, 60(15):4356--4365, May 2021.

\bibitem{luo2021metasurfaceenabled}
Xuhao Luo, Yueqiang Hu, Xin Li, Xiangnian Ou, Jiajie Lai, Na~Liu, and Huigao
  Duan.
\newblock Metasurface-{{Enabled On}}-{{Chip Multiplexed Diffractive Neural
  Networks}} in the {{Visible}}.
\newblock {\em arXiv:2107.07873 [physics]}, July 2021.

\bibitem{pu2020automatic}
Guoqing Pu, Li~Zhang, Weisheng Hu, and Lilin Yi.
\newblock Automatic mode-locking fiber lasers: Progress and perspectives.
\newblock {\em Science China Information Sciences}, 63(6):160404, May 2020.

\bibitem{singh2020mapping}
Robin Singh, Robin Singh, Robin Singh, Anu Agarwal, Anu Agarwal, Anu Agarwal,
  Brian~W. Anthony, Brian~W. Anthony, and Brian~W. Anthony.
\newblock Mapping the design space of photonic topological states via deep
  learning.
\newblock {\em Optics Express}, 28(19):27893--27902, September 2020.

\bibitem{wu2020machine}
Bei Wu, Kun Ding, C.T. Chan, and Yuntian Chen.
\newblock Machine {{Prediction}} of {{Topological Transitions}} in {{Photonic
  Crystals}}.
\newblock {\em Physical Review Applied}, 14(4):044032, October 2020.

\bibitem{ma2021universal}
Xuezhi Ma, Yuan Ma, Preston Cunha, Qiushi Liu, Kaushik Kudtarkar, Da~Xu, Jiafei
  Wang, Ming Liu, M.~Cynthia Hipwell, and Shoufeng Lan.
\newblock A universal deep learning strategy for designing high-quality-factor
  photonic resonances.
\newblock {\em arXiv:2105.03001 [physics]}, May 2021.

\bibitem{wang2021demonstration}
Zhedong Wang, Chao Qian, Tong Cai, Longwei Tian, Zhixiang Fan, Jian Liu, Yichen
  Shen, Li~Jing, Jianming Jin, Er-Ping Li, Bin Zheng, and Hongsheng Chen.
\newblock Demonstration of {{Spider}}-{{Eyes}}-{{Like Intelligent Antennas}}
  for {{Dynamically Perceiving Incoming Waves}}.
\newblock {\em Advanced Intelligent Systems}, n/a(n/a):2100066, July 2021.

\bibitem{wang2020topological}
Hongfei Wang, Samit~Kumar Gupta, Biye Xie, and Minghui Lu.
\newblock Topological photonic crystals: A review.
\newblock {\em Frontiers of Optoelectronics}, 13(1):50--72, March 2020.

\bibitem{segev2021topological}
Mordechai Segev and Miguel~A. Bandres.
\newblock Topological photonics: {{Where}} do we go from here?
\newblock {\em Nanophotonics}, 10(1):425--434, January 2021.

\bibitem{pilozzi2018machine}
Laura Pilozzi, Francis~A. Farrelly, Giulia Marcucci, and Claudio Conti.
\newblock Machine learning inverse problem for topological photonics.
\newblock {\em Communications Physics}, 1(1):1--7, September 2018.

\bibitem{long2019inverse}
Yang Long, Jie Ren, Yunhui Li, and Hong Chen.
\newblock Inverse design of photonic topological state via machine learning.
\newblock {\em Applied Physics Letters}, 114(18):181105, May 2019.

\bibitem{pilozzi2021topological}
Laura Pilozzi, Francis~A. Farrelly, Giulia Marcucci, and Claudio Conti.
\newblock Topological nanophotonics and artificial neural networks.
\newblock {\em Nanotechnology}, 32(14):142001, January 2021.

\bibitem{chen2020inversea}
Yafeng Chen, Fei Meng, Yuri Kivshar, Baohua Jia, and Xiaodong Huang.
\newblock Inverse design of higher-order photonic topological insulators.
\newblock {\em Physical Review Research}, 2(2):023115, May 2020.

\bibitem{nussbaum2021inverse}
Eric Nussbaum, Erik Sauer, and Stephen Hughes.
\newblock Inverse design of broadband and lossless topological photonic crystal
  waveguide modes.
\newblock {\em Optics Letters}, 46(7):1732--1735, April 2021.

\bibitem{zhirihin2021topological}
Dmitry~V. Zhirihin and Yuri~S. Kivshar.
\newblock Topological {{Photonics}} on a {{Small Scale}}.
\newblock {\em arXiv:2108.12248 [cond-mat, physics:physics]}, August 2021.

\bibitem{koshelev2019metaoptics}
Kirill Koshelev, Andrey Bogdanov, and Yuri Kivshar.
\newblock Meta-optics and bound states in the continuum.
\newblock {\em Science Bulletin}, 64(12):836–842, Jun 2019.

\bibitem{lin2021engineering}
Ronghui Lin, Zahrah Alnakhli, and Xiaohang Li.
\newblock Engineering of multiple bound states in the continuum by latent
  representation of freeform structures.
\newblock {\em Photonics Research}, 9(4):B96--B103, April 2021.

\bibitem{azzam2021photonic}
Shaimaa~I. Azzam and Alexander~V. Kildishev.
\newblock Photonic bound states in the continuum: From basics to applications.
\newblock {\em Advanced Optical Materials}, 9(1):2001469, 2021.

\bibitem{sadreev2021interference}
Almas~F Sadreev.
\newblock Interference traps waves in an open system: bound states in the
  continuum.
\newblock {\em Reports on Progress in Physics}, 84(5):055901, apr 2021.

\bibitem{shastri_photonics_2021}
Bhavin~J. Shastri, Alexander~N. Tait, T.~Ferreira~de Lima, Wolfram H.~P.
  Pernice, Harish Bhaskaran, C.~D. Wright, and Paul~R. Prucnal.
\newblock Photonics for artificial intelligence and neuromorphic computing.
\newblock {\em Nature Photonics}, 15(2):102--114, February 2021.
\newblock Bandiera\_abtest: a Cg\_type: Nature Research Journals Number: 2
  Primary\_atype: Reviews Publisher: Nature Publishing Group Subject\_term:
  Integrated optics;Nanophotonics and plasmonics;Optoelectronic devices and
  components;Ultrafast photonics Subject\_term\_id:
  integrated-optics;nanophotonics-and-plasmonics;optoelectronic-devices-and-components;ultrafast-photonics.

\bibitem{schuman_survey_2017}
Catherine~D. Schuman, Thomas~E. Potok, Robert~M. Patton, J.~Douglas Birdwell,
  Mark~E. Dean, Garrett~S. Rose, and James~S. Plank.
\newblock A {Survey} of {Neuromorphic} {Computing} and {Neural} {Networks} in
  {Hardware}.
\newblock May 2017.

\bibitem{shen_deep_2017}
Yichen Shen, Nicholas~C. Harris, Scott Skirlo, Mihika Prabhu, Tom Baehr-Jones,
  Michael Hochberg, Xin Sun, Shijie Zhao, Hugo Larochelle, Dirk Englund, and
  Marin Soljačić.
\newblock Deep learning with coherent nanophotonic circuits.
\newblock {\em Nature Photonics}, 11(7):441--446, July 2017.
\newblock Bandiera\_abtest: a Cg\_type: Nature Research Journals Number: 7
  Primary\_atype: Research Publisher: Nature Publishing Group Subject\_term:
  Integrated optics;Silicon photonics Subject\_term\_id:
  integrated-optics;silicon-photonics.

\bibitem{hughes2018training}
Tyler~W. Hughes, Momchil Minkov, Yu~Shi, and Shanhui Fan.
\newblock Training of photonic neural networks through in situ backpropagation
  and gradient measurement.
\newblock {\em Optica}, 5(7):864--871, July 2018.

\bibitem{feldmann_all-optical_2019}
J.~Feldmann, N.~Youngblood, C.~D. Wright, H.~Bhaskaran, and W.~H.~P. Pernice.
\newblock All-optical spiking neurosynaptic networks with self-learning
  capabilities.
\newblock {\em Nature}, 569(7755):208--214, May 2019.
\newblock Bandiera\_abtest: a Cg\_type: Nature Research Journals Number: 7755
  Primary\_atype: Research Publisher: Nature Publishing Group Subject\_term:
  Microresonators;Nanophotonics and plasmonics Subject\_term\_id:
  microresonators;nanophotonics-and-plasmonics.

\bibitem{lin_all-optical_2018}
Xing Lin, Yair Rivenson, Nezih~T. Yardimci, Muhammed Veli, Yi~Luo, Mona
  Jarrahi, and Aydogan Ozcan.
\newblock All-optical machine learning using diffractive deep neural networks.
\newblock {\em Science}, 361(6406):1004--1008, September 2018.
\newblock Publisher: American Association for the Advancement of Science.

\bibitem{wright_deep_2021}
Logan~G. Wright, Logan~G. Wright, Tatsuhiro Onodera, Tatsuhiro Onodera,
  Martin~M. Stein, Tianyu Wang, Darren~T. Schachter, Zoey Hu, and Peter~L.
  McMahon.
\newblock Deep nonlinear optical neural networks using physics-aware training.
\newblock In {\em Conference on {Lasers} and {Electro}-{Optics} (2021), paper
  {FF1A}.4}, page FF1A.4. Optical Society of America, May 2021.

\bibitem{majumdar_metaphotonic_2020}
Arka Majumdar and Arka Majumdar.
\newblock Metaphotonic {Computational} {Image} {Sensors}.
\newblock In {\em Imaging and {Applied} {Optics} {Congress} (2020), paper
  {IW1D}.3}, page IW1D.3. Optical Society of America, June 2020.

\bibitem{specht_brief_nodate}
Donald~F. Specht.
\newblock Brief {Papers} {A} {General} {Regression} {Neural} {Network}.

\bibitem{lee2021retinainspired}
Kyuho Lee, Hyowon Han, Youngwoo Kim, Jumi Park, Seonghoon Jang, Hyeokjung Lee,
  Seung~Won Lee, HoYeon Kim, Yeeun Kim, Taebin Kim, Dongho Kim, Gunuk Wang, and
  Cheolmin Park.
\newblock Retina-{{Inspired Structurally Tunable Synaptic Perovskite
  Nanocones}}.
\newblock {\em Advanced Functional Materials}, n/a(n/a):2105596, August 2021.

\bibitem{hu2020singlenanoparticle}
Jingtian Hu, Tingting Liu, Priscilla Choo, Shengjie Wang, Thaddeus Reese,
  Alexander~D. Sample, and Teri~W. Odom.
\newblock Single-{{Nanoparticle Orientation Sensing}} by {{Deep Learning}}.
\newblock {\em ACS Central Science}, 6(12):2339--2346, December 2020.

\bibitem{shiratori2021machinelearned}
Katsuya Shiratori, Logan D.~C. Bishop, Behnaz Ostovar, Rashad Baiyasi, Yi-Yu
  Cai, Peter~J. Rossky, Christy~F. Landes, and Stephan Link.
\newblock Machine-{{Learned Decision Trees}} for {{Predicting Gold Nanorod
  Sizes}} from {{Spectra}}.
\newblock {\em The Journal of Physical Chemistry C}, August 2021.

\bibitem{pu2021unlabeled}
Tanchao Pu, Jun-Yu Ou, Vassili Savinov, Guanghui Yuan, Nikitas Papasimakis, and
  Nikolay~I. Zheludev.
\newblock Unlabeled {{Far}}-{{Field Deeply Subwavelength Topological
  Microscopy}} ({{DSTM}}).
\newblock {\em Advanced Science}, 8(1):2002886, 2021.

\bibitem{shao2020machine}
Siyao Shao, Siyao Shao, Kevin Mallery, Kevin Mallery, S.~Santosh Kumar,
  S.~Santosh Kumar, Jiarong Hong, and Jiarong Hong.
\newblock Machine learning holography for {{3D}} particle field imaging.
\newblock {\em Optics Express}, 28(3):2987--2999, February 2020.

\bibitem{wang2021deeplearningassisted}
Qian Wang, Hua He, Qian Zhang, Zhenzhen Feng, Jiqiang Li, Xiaoliang Chen, Lihua
  Liu, Xiaojuan Wang, Baosheng Ge, Daoyong Yu, Hao Ren, and Fang Huang.
\newblock Deep-{{Learning}}-{{Assisted Single}}-{{Molecule Tracking}} on a
  {{Live Cell Membrane}}.
\newblock {\em Analytical Chemistry}, 93(25):8810--8816, June 2021.

\bibitem{speiser2021deep}
Artur Speiser, Lucas-Raphael M{\"u}ller, Philipp Hoess, Ulf Matti,
  Christopher~J. Obara, Wesley~R. Legant, Anna Kreshuk, Jakob~H. Macke, Jonas
  Ries, and Srinivas~C. Turaga.
\newblock Deep learning enables fast and dense single-molecule localization
  with high accuracy.
\newblock {\em Nature Methods}, pages 1--9, September 2021.

\bibitem{zhang2018analyzing}
Peiyi Zhang, Sheng Liu, Abhishek Chaurasia, Donghan Ma, Michael~J.
  Mlodzianoski, Eugenio Culurciello, and Fang Huang.
\newblock Analyzing complex single-molecule emission patterns with deep
  learning.
\newblock {\em Nature Methods}, 15(11):913--916, November 2018.

\bibitem{tranter_multiparameter_2018}
A.~D. Tranter, H.~J. Slatyer, M.~R. Hush, A.~C. Leung, J.~L. Everett, K.~V.
  Paul, P.~Vernaz-Gris, P.~K. Lam, B.~C. Buchler, and G.~T. Campbell.
\newblock Multiparameter optimisation of a magneto-optical trap using deep
  learning.
\newblock {\em Nature Communications}, 9(1):4360, October 2018.
\newblock Bandiera\_abtest: a Cc\_license\_type: cc\_by Cg\_type: Nature
  Research Journals Number: 1 Primary\_atype: Research Publisher: Nature
  Publishing Group Subject\_term: Atom optics;Computational science;Quantum
  information;Ultracold gases Subject\_term\_id:
  atom-optics;computational-science;quantum-information;ultracold-gases.

\bibitem{gupta2021machine}
Ratnesh~K. Gupta, Jesse~L. Everett, Aaron~D. Tranter, René Henke, Vandna
  Gokhroo, Ping~Koy Lam, and Síle~Nic Chormaic.
\newblock Machine learner optimization of optical nanofiber-based dipole traps
  for cold $^{87}$rb atoms, 2021.

\bibitem{wigley_fast_2016}
P.~B. Wigley, P.~J. Everitt, A.~van~den Hengel, J.~W. Bastian, M.~A.
  Sooriyabandara, G.~D. McDonald, K.~S. Hardman, C.~D. Quinlivan, P.~Manju,
  C.~C.~N. Kuhn, I.~R. Petersen, A.~N. Luiten, J.~J. Hope, N.~P. Robins, and
  M.~R. Hush.
\newblock Fast machine-learning online optimization of ultra-cold-atom
  experiments.
\newblock {\em Scientific Reports}, 6:srep25890, May 2016.

\bibitem{henson_approaching_2018}
Bryce~M. Henson, Dong~K. Shin, Kieran~F. Thomas, Jacob~A. Ross, Michael~R.
  Hush, Sean~S. Hodgman, and Andrew~G. Truscott.
\newblock Approaching the adiabatic timescale with machine learning.
\newblock {\em Proceedings of the National Academy of Sciences},
  115(52):13216--13221, December 2018.
\newblock Publisher: National Academy of Sciences Section: Physical Sciences.

\end{thebibliography}

\end{document}